\documentclass[amsmath, amssymb, floatfix, reprint, twocolumn, notitlepage, 10pt]{revtex4-1}
\usepackage[T1]{fontenc}
\usepackage[utf8]{inputenc}
\usepackage{microtype,bm,bbm,graphicx,booktabs,times}
\usepackage[usenames,dvipsnames]{xcolor}
\usepackage{setspace}
\selectfont{}

\usepackage{hyperref}
\usepackage{braket}
\usepackage{subfiles} 
\usepackage{upgreek}
\hypersetup{colorlinks,allcolors=black,linktoc=all}
\usepackage[capitalize]{cleveref}
\crefformat{section}{Sec.~#2#1#3}

\begin{document}
\title{Optimal two-photon excitation of bound states in non-Markovian waveguide QED}
\author{Rahul Trivedi$^{1, 2}$}
\email{rtrivedi@stanford.edu}
\author{Daniel Malz$^{3, 4}$}
\author{Shanhui Fan$^{1,2}$}
\author{Jelena Vu\v{c}kovi\'c$^{1,2}$}
\affiliation{{$^1$E. L. Ginzton Laboratory, Stanford University, Stanford, CA 94305, USA. \\
	         $^2$Department of Electrical Engineering, Stanford, CA 94305, USA.\\
	         $^3$Max-Planck-Institute for Quantum Optics, Hans-Kopfermann-Str.~1, 85748 Garching, Germany.\\
	         $^4$Munich Center for Quantum Science and Technology (MCQST), Schellingstra{\ss}e 4, 80799 M{\"u}nchen, Germany.}}

\date{\today}

\begin{abstract}
Bound states arise in waveguide QED systems with a strong frequency-dependence of the coupling between emitters and photonic modes. While exciting such bound-states with single photon wave-packets is not possible, photon-photon interactions mediated by the emitters can be used to excite them with two-photon states. In this letter, we use scattering theory to provide upper limits on this excitation probability for a general non-Markovian waveguide QED system and show that this limit can be reached by a two-photon wave-packet with vanishing uncertainty in the total photon energy. Furthermore, we also analyze multi-emitter waveguide QED systems with multiple bound states and provide a systematic construction of two-photon wave-packets that can excite a given superposition of these bound states. As specific examples, we study bound state trapping in waveguide QED systems with single and multiple emitters and a time-delayed feedback.
\end{abstract}
\maketitle

\noindent\emph{Introduction}: Waveguide quantum electrodynamics (wQED) \cite{shen2007strongly, shen2005coherent, gonzalez2014generation, shi2013two, rephaeli2012few} studies the interaction of quantum emitters with one-dimensional bosonic waveguide fields. While traditional analysis of wQED systems assumes a Markovian (frequency-independent) coupling of emitters and the waveguide mode \cite{xu2013analytic, xu2015input, xu2017input, trivedi2018few, caneva2015quantum}, there has been recent theoretical interest in exploring physics of non-Markovian wQED systems \cite{dinc2019exact, fang2015waveguide, shi2015multiphoton, pichler2016photonic, gonzalez2017quantum, calajo2016atom,calajo2019exciting, carmele2020pronounced, sinha2020non, dinc2019non, sinha2020collective, kockum2018decoherence, mirza2016multiqubit, zheng2013persistent}. In particular, several recent works have attempted to understand the dynamics of wQED systems with time-delays comparable to or larger than the lifetime of the emitters. Such non-Markovian wQED systems support a rich variety of physical phenomena including existence of bound states in continuum \cite{calajo2016atom,calajo2019exciting}, superradiance and subradiance in the presence of time delays \cite{carmele2020pronounced, sinha2020non, dinc2019non, sinha2020collective} as well as generation of highly entangled photonic states \cite{ramos2016non, pichler2017universal, mirza2016multiqubit, zheng2013persistent}. Furthermore, there is a possibility of using these physical phenomena for quantum technology applications such as quantum memory \cite{calajo2019exciting} and quantum computation with cluster states \cite{pichler2017universal}. 

Of particular interest in non-Markovian wQED is the existence of single-excitation polaritonic bound states \cite{calajo2016atom,calajo2019exciting}, which are normalizable eigenstates of the wQED Hamiltonian. The energy of these bound states lies within the continuum of frequencies supported by the waveguide mode, thus opening up the possibility of exciting them efficiently through the waveguide. While these bound states cannot be excited with single waveguide photons, the emitter-mediated photon-photon interactions can allow two (or more) waveguide photons to excite them \cite{calajo2019exciting}. From a technological standpoint, this opens up the possibility of storing quantum information being carried by two-photon wave-packets into the bound states. To this end, Ref.~\cite{calajo2019exciting} computationally studied the two-photon excitation of the bound state in a wQED system with an emitter and time-delayed feedback and achieved $\sim$85$\%$ bound state trapping probability by designing the two-photon wave-packet. However, it remains unclear what the limits on bound state trapping probabilities are, and if there is a systematic design procedure for the optimal incident two-photon wave-packet that reaches this limit.

In this letter, we use quantum scattering theory to rigorously answer this question. Our approach relies on re-expressing the wQED Hamiltonian in terms of its bound states and scattering states coupled to each other via the anharmonicity of the emitters and analytically calculating the two-photon scattering matrix element capturing the bound state trapping process. Using this scattering matrix, we provide an upper limit on the bound state trapping probability. Furthermore, we show that this limit is reached by a two-photon wave-packet with vanishing uncertainty in the total photon energy. Finally, as storage protocols for quantum information encoded in the incoming two-photon wave-packets, we consider multi-emitter wQED systems that can support more than 1 bound states and systematically outline the design of two-photon wave-packets to excite superpositions of these bound states.
\\ \ \\

\begin{figure}
\centering
\includegraphics[scale=0.6]{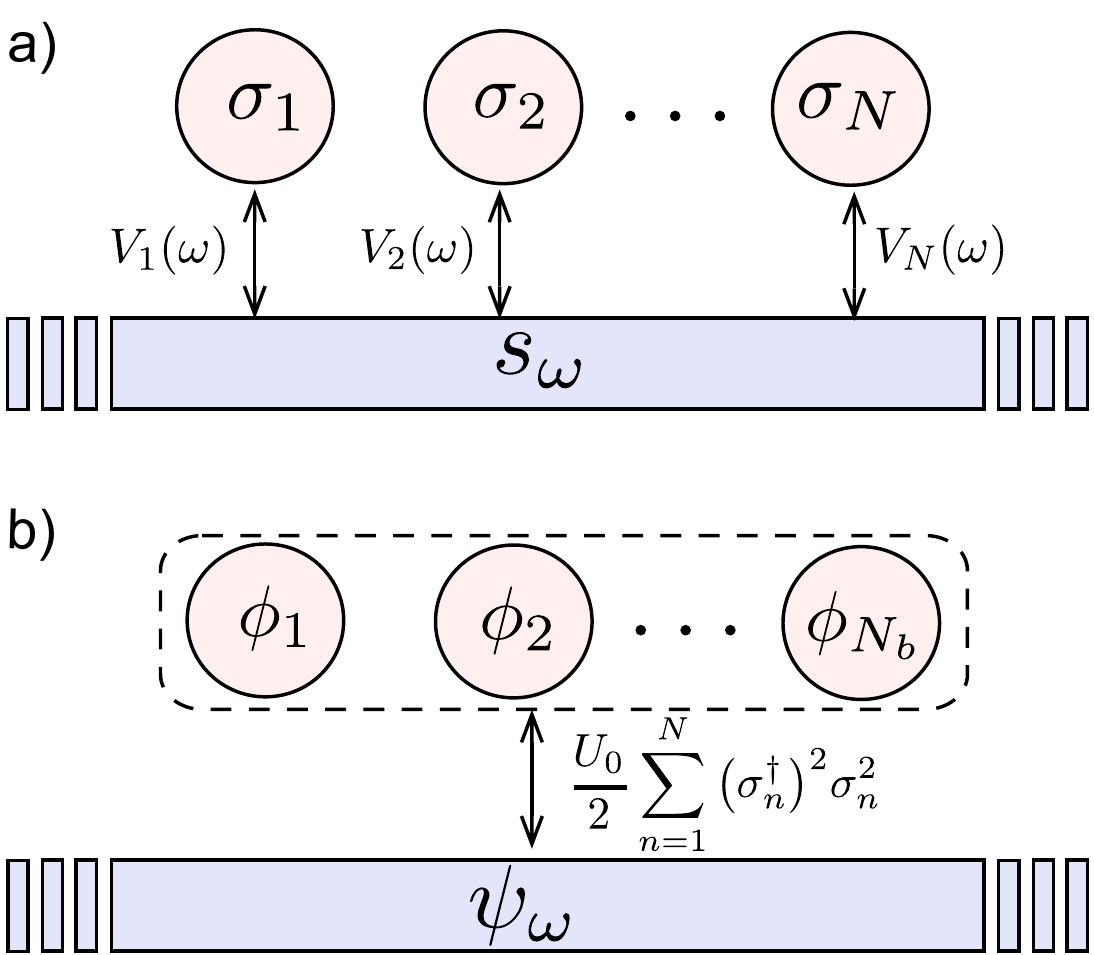}
\caption{a) Schematic of the non-Markovian waveguide QED system with $N$ emitters. The frequency-dependent coupling constant $V_n(\omega)$ capture the non-Markovian nature of the emitter-waveguide interactions. b) An equivalent picture of the waveguide QED system when expressed in terms of the scattering state modes and the bound state modes which are coupled to each other due to the two-particle repulsion at the qubit modes. }
\label{fig:schematic}
\end{figure}

\noindent\emph{Scattering Theory}: The wQED system under consideration is shown in Fig.~\ref{fig:schematic}a ---  $N$ emitters modeled as anharmonic oscillators at frequencies $\omega_1, \omega_2 \dots \omega_N$ with annihilation operators $\sigma_1, \sigma_2 \dots \sigma_N$ couple with coupling constant $V_1(\omega), V_2(\omega) \dots V_N(\omega)$ to a waveguide mode with annihilation operator $s_\omega$. The dynamics of this system can be described by a Hamiltonian expressible as $H = H_0 + V$ where $H_0$ is a quadratic form that describes the interaction of the emitters with the waveguide:
\begin{align}
H_0 =& \int_{-\infty}^\infty \omega s_\omega^\dagger s_\omega d\omega + \sum_{n=1}^N \omega_n \sigma_n^\dagger \sigma_n  \nonumber\\
& +\int_{-\infty}^\infty \sum_{n=1}^N\big(V_n(\omega) s_\omega \sigma_n^\dagger + V_n^*(\omega) \sigma_n s_\omega^\dagger\big) \frac{d\omega}{\sqrt{2\pi}},
\end{align}
and $V$ captures the anharmonicity of the emitters:
\begin{align}\label{eq:Hamiltonian}
V = \sum_{n=1}^N\frac{U_0}{2}\big(\sigma_n^\dagger\big)^2 \sigma_n^2.
\end{align}
It can be noted that two-level emitters are obtained in the limit of infinite anharmonicity ($U_0 \to \infty$).

The quadratic Hamiltonian $H_0$ can be diagonalized into the sum of a continuum of \emph{scattering states} with annihilation operators $\psi_\omega$ at frequencies $\omega \in \mathbb{R}$ and discrete \emph{bound states} with annihilation operators $\phi_1, \phi_2 \dots \phi_{N_b}$ at frequencies $\omega_1, \omega_2 \dots \omega_{N_b}$ (Fig.~\ref{fig:schematic}b):
\begin{align}\label{eq:quad_hamil_diag}
H_0 = \sum_{\alpha=1}^{N_b} \omega_\alpha \phi_\alpha^\dagger \phi_\alpha +  \int_{-\infty}^\infty \omega \psi_\omega^\dagger \psi_\omega d\omega.
\end{align}
The scattering state modes and the bound state modes, while decoupled in the hamiltonian $H_0$, are coupled due to the anharmonicity of the emitters (Eq.~\ref{eq:Hamiltonian}). Furthermore, the annihilation operators $\sigma_n$ for the emitters can be expressed in terms of the bound state operator and scattering state operators:
\begin{align}\label{eq:olap_def}
\sigma_n = \sum_{\alpha = 1}^{N_b}\varepsilon^\alpha_n \phi_\alpha + \int_{-\infty}^\infty \xi_n(\omega) \psi_\omega d\omega,
\end{align}
where $\varepsilon_n^\alpha$ captures the overlap of the $\alpha^\text{th}$ bound state mode with the $n^\text{th}$ emitter and $\xi_n(\omega)$ captures the overlap of the scattering state mode at frequency $\omega$ with the $n^\text{th}$ emitter. The diagonalization of the Hamiltonian as well as the computation of overlap of the emitter modes with the bound state modes and the scattering state modes is discussed in the supplement \cite{supp}.

\begin{figure*}[htpb]
\includegraphics[scale=0.45]{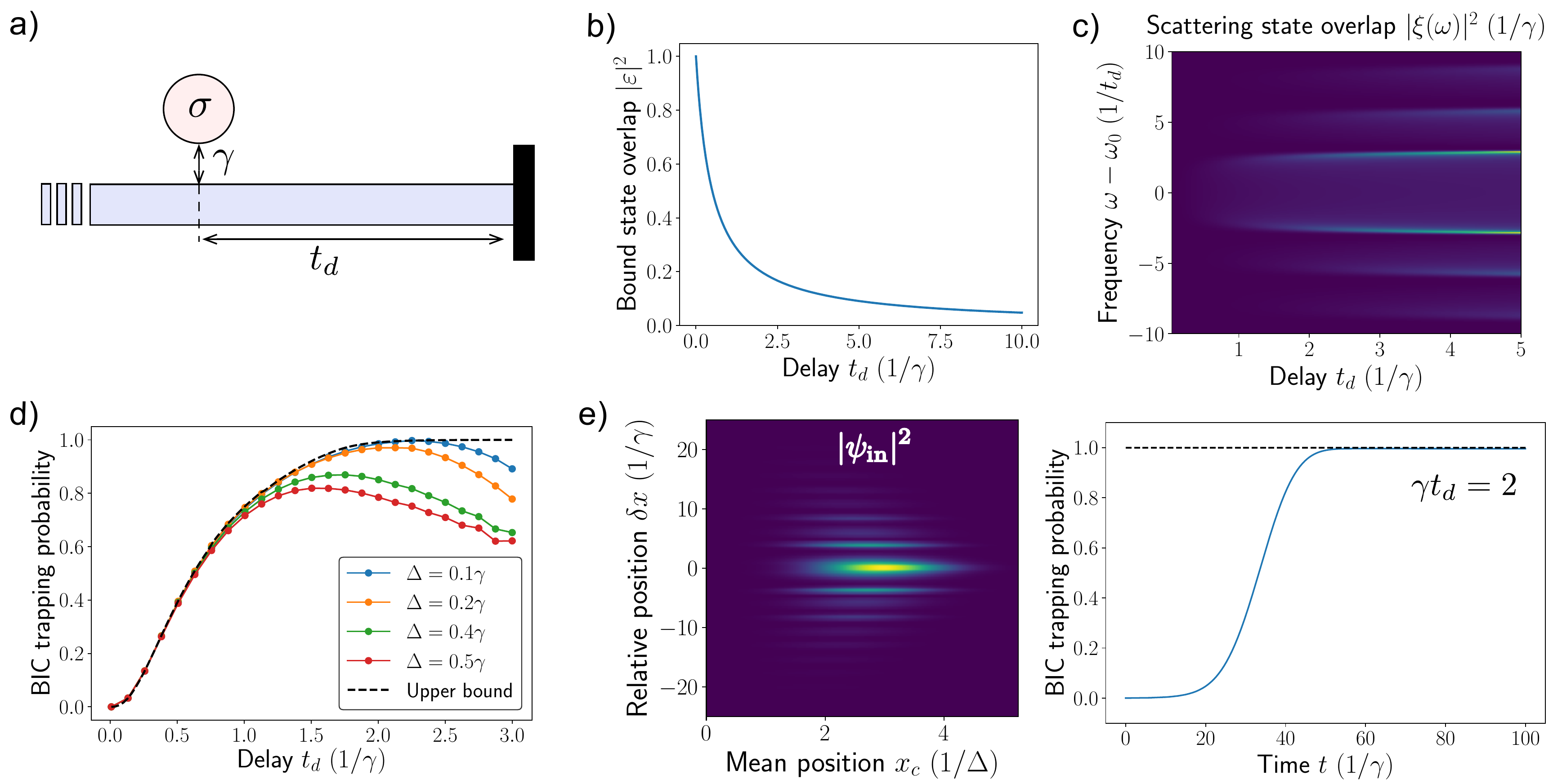}
\caption{\textbf{Optimal excitation of bound states} in time-delayed feedback system. a) Schematic of a time-delayed feedback system with a single emitter connected to a waveguide mode terminated by a mirror. b) Overlap of bound state with the emitter as a function of time delay $t_d$. c) Overlap of the scattering state at frequency $\omega$ with the emitter for different time delays $t_d$. d) Upper bound on the two-photon excitation probability (dashed black line) as a function of the delay as well as the probability obtained on using two-photon wave-packets for different uncertainties in the two-photon energies. d) Finite-difference time-domain simulations of the time-delayed feedback system with the incident two-photon state showed as a function of photon positions. It can be seen that the bound state is excited with nearly unity trapping probability with this incident two-photon state. The incident two-photon state is constructed from Eq.~\ref{eq:inc_two_ph_single_bs} with central two-photon energy $\Omega_0 = 2\omega_0 + 0.95\gamma$ and energy uncertainty $\Delta=0.1\gamma$.}
\label{fig:single_bs_ex}
\end{figure*}

Consider now the process of exciting the emitters with an incident two-photon state and trapping one photon in a bound state. The probability amplitude associated with this process is captured by the following scattering matrix element:
\begin{align}\label{eq:smat_def}
&S_\alpha(\omega; \nu_1, \nu_2) = \lim_{\substack{t_i\to-\infty \\ t_f \to \infty}}\bra{\text{G}} \phi_\alpha \psi_\omega U_I(t_f, t_i)\psi_{\nu_1}^\dagger \psi_{\nu_2}^\dagger \ket{\text{G}},
\end{align}
where $U_I(\cdot, \cdot)$ is the interaction picture propagator for the Hamiltonian $H$ with respect to $H_0$ and $\ket{\text{G}}$ is the ground state of the wQED system. $S_\alpha(\omega; \nu_1, \nu_2)$ is the probability amplitude of trapping a photon in the $\alpha^\text{th}$ bound-state and scattering the second photon in a scattering state at frequency $\omega$ on excitation with two photons at frequency $\nu_1$ and $\nu_2$. As is shown in the supplement \cite{supp}, an analytical expression relating this scattering matrix to $\varepsilon_n^\alpha$ and $\xi_n(\omega)$ can be derived by following a procedure similar to Ref.~\cite{shi2015multiphoton}: The propagator in Eq.~\ref{eq:smat_def} can be expanded into a Dyson series, with each term in the Dyson series being evaluated using the Wick's theorem. The resulting series can then be analytically summed to obtain
\begin{subequations}
\begin{align}\label{eq:smat_elem_final}
S_\alpha(\omega; \nu_1, \nu_2) = \Gamma_\alpha(\omega; \nu_1, \nu_2)\delta(\omega + \omega_\alpha - \nu_1 - \nu_2),
\end{align}
where in the limit of $U_0 \to \infty$,
\begin{align}\label{eq:smat_conn_anh}
\Gamma_\alpha(\omega; \nu_1, \nu_2) &= -4\pi \sum_{m,n = 1}^{N}\bigg(\varepsilon_m^{\alpha*} \xi_m^*(\omega)\times \nonumber\\ &\big[\textbf{T}^{-1}(\omega + \omega_\alpha + i0^+)\big]_{m, n} \xi_{n}(\nu_1) \xi_n(\nu_2)\bigg).
\end{align}
Here $\textbf{T}(\Omega + i0^+)$ is a $N \times N$ matrix defined by
\begin{align}
\big[\textbf{T}(\Omega + i0^+)\big]_{m, n} = \int_0^\infty G_{m, n}^2(t)e^{i\Omega t} dt,
\end{align}
where
\begin{align}
G_{m, n}(t) = \sum_{n=1}^{N_b}\varepsilon_m^\alpha \varepsilon_n^{\alpha*} e^{-i\omega_\alpha t} + \int_{-\infty}^\infty \xi_m(\omega)\xi_n^*(\omega) e^{-i\omega t}d\omega.
\end{align}
\end{subequations}
The delta function singularity in Eq.~\ref{eq:smat_elem_final} constrains the output photon frequency $\omega$ given input photon frequencies $\nu_1$ and $\nu_2$ as required by energy conservation. Furthermore, the matrix $\textbf{T}(\Omega + i0^+)$ captures the two-excitation dynamics of the multi-emitter wQED system. Finally, Eq.~\ref{eq:smat_conn_anh} relate the scattering amplitude $\Gamma_\alpha(\omega; \nu_1, \nu_2)$ to this matrix and the overlap of the bound states and scattering states with the emitters.

\noindent \emph{Optimal trapping probability}: We now consider exciting the system with a two-photon state described by a wavefunction $\psi_\text{in}(\nu_1, \nu_2)$:
\begin{align}
 \ket{\psi_\text{in}} = \frac{1}{\sqrt{2}} \int_{\nu_1, \nu_2 =-\infty}^\infty \psi_\text{in}(\nu_1, \nu_2)\psi_{\nu_1}^\dagger \psi_{\nu_2}^\dagger \ket{\text{vac}}d\nu_1 d\nu_2
\end{align}
Using the scattering matrix element in Eq.~\ref{eq:smat_elem_final}, we can obtain the bound state trapping probability:
\begin{align}\label{eq:trap_prob}
&P_\alpha[\psi_\text{in}] =\nonumber\\ &\frac{1}{2} \int_{-\infty}^\infty d\Omega \bigg|\int_{-\infty}^\infty \Gamma_\alpha(\Omega - \omega_\alpha; \nu,\Omega - \nu) \psi_\text{in}(\nu,\Omega - \nu) d\nu \bigg|^2
\end{align}
This result allows us to upper bound the trapping probability. As is shown in the supplement \cite{supp}, a direct application of the Cauchy Schwarz inequality yields $P_\alpha[\psi_\text{in}] < P_{\alpha}^{\text{ub}}$ where:
\begin{align}\label{eq:ineq_single_bs}
P_\alpha^{\text{ub}} =  \max_{\Omega\in \mathbb{R}}\bigg(\frac{1}{2} \int_{-\infty}^\infty  d\nu \big|\Gamma_\alpha(\Omega - \omega_\alpha; \nu, \Omega - \nu)\big|^2 \bigg).
\end{align}
Furthermore, it follows from Eq.~\ref{eq:trap_prob} that an energy entangled two-photon wave packet can get arbitrarily close to this bound provided that the uncertainty in the total photon energy is sufficiently small. More specifically, consider a family of photon wave-packets $\psi_{\alpha, \Delta}(\nu_1, \nu_2)$ defined by
\begin{align}\label{eq:inc_two_ph_single_bs}
\psi_{\alpha, \Delta}&(\nu_1, \nu_2) = N_{\alpha, \Delta} f_{\Delta, \Omega_0}(\nu_1 + \nu_2) \Gamma^*_\alpha(\Omega_0 - \omega_\alpha; \nu_1, \nu_2)
\end{align}
where $f_{\Delta, \Omega_0}(\nu) = (\pi\Delta^2)^{-1/4} \exp(-(\nu - \Omega_0)^2 / 2\Delta^2)$ determines the distribution of two-photon energy, $\Omega_0$ is the central two-photon energy chosen as the frequency that maximizes the right hand side of Eq.~\ref{eq:ineq_single_bs} and $N_{\alpha, \Delta}$ is chosen to normalize the wave-packet. It then follows from Eq.~\ref{eq:trap_prob} that as $\Delta \to 0$, $P_\alpha[\psi_{\alpha, \Delta}] \to P_\alpha^\text{ub}$.

\begin{figure*}[htpb]
\centering
\includegraphics[scale=0.33]{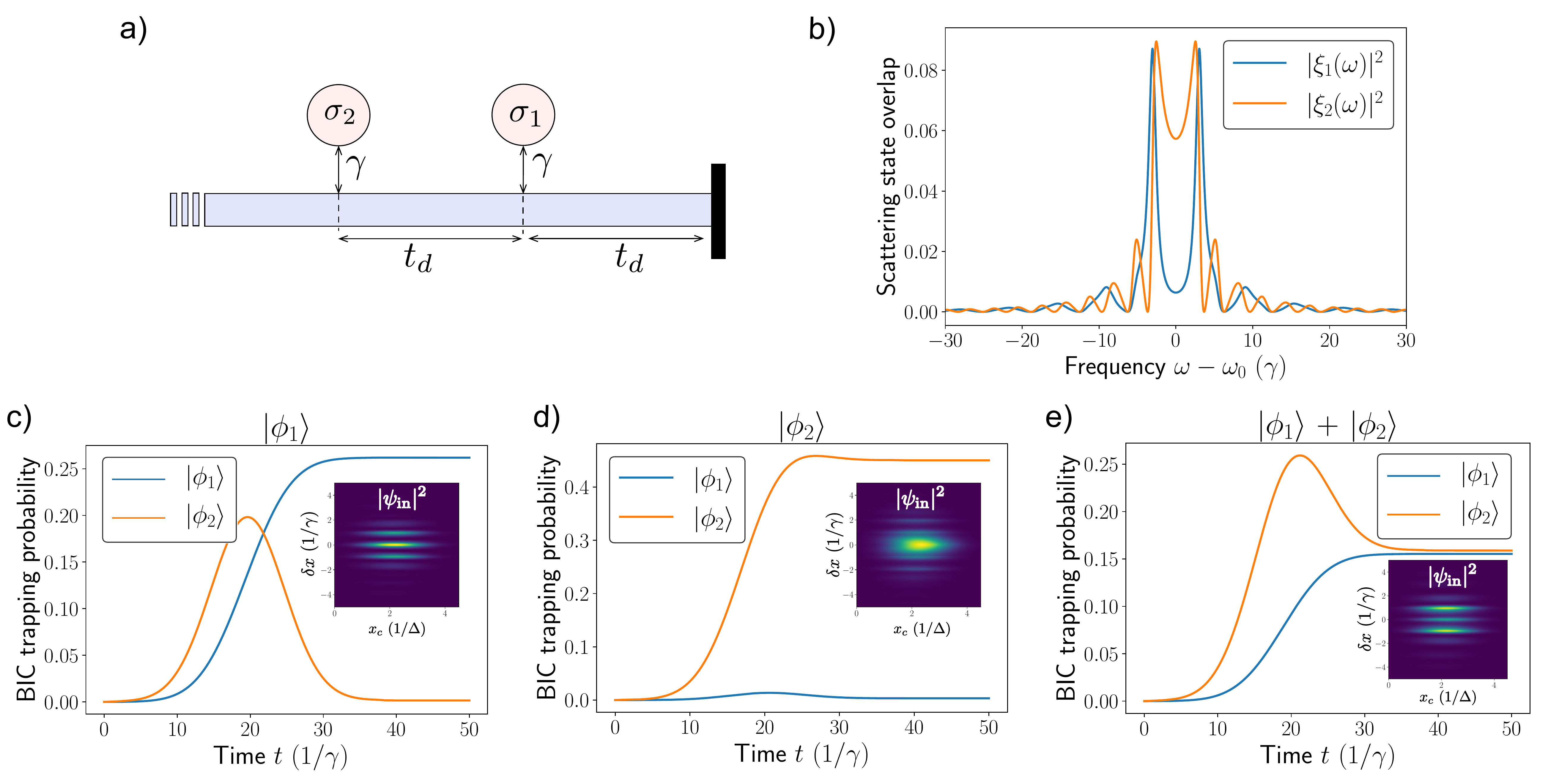}
\caption{\textbf{Excitation of bound state superpositions} in time-delayed feedback system. a) Schematic of a time-delayed feedback system with two emitters connected to a waveguide mode terminated by a mirror. b) Overlap of the scattering states with the two emitters as a function of the scattering state mode frequency.  Finite-difference time-domain simulations showing trapping c) first bound state, d) second bound state and d) equal superposition of the two with appropriately chosen two-photon wave-packets (shown as insets). In all the simulations, the time-delay $t_d$ is assumed to be $0.5 / \gamma$, the central frequency $\Omega_0$ of the wave-packet is chosen to be $2.4\gamma$ and the two-photon energy uncertainty $\Delta$ is chosen to be $0.15\gamma$.}
\label{fig:two_bs_ex}
\end{figure*}

As a concrete example, we consider a wQED system with time-delayed feedback as shown in Fig.~\ref{fig:single_bs_ex}a. This system is equivalent to a non-Markovian waveguide QED system with one emitter and $V(\omega) = 2i \sqrt{\gamma}\sin(\omega t_d)$. If the qubit transition frequency $\omega_0$ satisfies $\omega_0 t_d = n\pi$ for some integer $n$, then this system supports one bound state mode. Furthermore, the overlap of the qubit mode with the bound state ($\varepsilon$) and the scattering state $\big(\xi(\omega)\big)$ can be computed by diagonalizing the quadratic part of the system Hamiltonian (refer to supplement \cite{supp} for details):
\begin{align}\label{eq:tdf_olaps}
\varepsilon = \frac{1}{\sqrt{1 + 2\gamma t_d}} \ \text{and} \ \xi(\omega) = \frac{2i\sqrt{\gamma}\sin(\omega t_d)}{\omega - \omega_0 + 2\gamma\sin(\omega t_d)e^{-i\omega t_d}}
\end{align}
In the short delay regime ($\gamma t_d \ll 1$) the bound state is completely localized to the emitter (Fig.~\ref{fig:single_bs_ex}b). Furthermore, in this regime the system is Markovian with vanishing coupling between the emitter and waveguide ($V(\omega)\approx V(\omega_0) = 0$) and consequently the bound state trapping probability vanishes (Fig.~\ref{fig:single_bs_ex}d). In the long-delay regime ($\gamma t_d \gg 1$), the overlap of the emitter with the scattering states becomes significant (Fig.~\ref{fig:single_bs_ex}c). We find that in this regime, the upper bound on the trapping probability reaches 1 (Fig.~\ref{fig:single_bs_ex}d). However, the energy uncertainty $\Delta$ needed to reach this bound decreases with an increase in the delay $t_d$ (Fig.~\ref{fig:single_bs_ex}c) implying that the incident wave-packet is increasingly unconfined in space. Figure \ref{fig:single_bs_ex}e shows Finite Difference Time Domain (FDTD) simulation of two-photon scattering  \cite{fang2019fdtd, supp} from this system for $\gamma t_d = 2$, and we indeed see that the bound state can be excited with near unity probability consistent with the scattering theory results.

\noindent \emph{Exciting bound-state superpositions}: Multi-emitter non-Markovian wQED systems can support more than one single-excitation bound states. An incident two-photon wave-packet will, in general, excite a superposition of bound states that is controllable by engineering the two-photon wave-packet. This opens up the possibility of using such systems for large quantum memories, with the number of bound states determining the size of the quantum memory.

Since the scattering amplitude in Eq.~\ref{eq:smat_conn_anh} suggests that the superposition of the bound states being excited depends on the overlap of $\psi_\text{in}(\nu_1, \nu_2)$ with $\xi_n(\nu)$, we assume the following ansatz for $\psi_\text{in}(\nu_1, \nu_2)$:
\begin{align}\label{eq:inc_two_ph_two_bs}
\psi_\text{in}(\nu_1, \nu_2) = f_{\Delta, \Omega_0}(\nu_1 + \nu_2) \sum_{n=1}^N c_n^{\text{in}} \xi_n^*(\nu_1) \xi_n^*(\nu_2),
\end{align}
where $f_{\Delta, \Omega_0}(\nu) = (\pi\Delta^2)^{-1/4}\exp(-(\nu - \Omega_0)^2 / 2\Delta^2)$ determines the distribution of the two-photon energy and the coefficients $c_n^\text{in}$ specify the spectral distribution of the two-photons. Under the assumption of negligble energy uncertainty ($\Delta \to 0$), an application of the scattering matrix in Eq.~\ref{eq:smat_elem_final} yields the following state:
\begin{align}\label{eq:out_state}
\ket{\psi_\text{out}} = \sum_{\alpha=1}^{N_b} c_\alpha^{\text{out}} \int_{-\infty}^\infty f_{\Delta, \Omega_0}(\omega + \omega_\alpha)\psi_{\omega}^\dagger \phi_\alpha^\dagger \ket{\text{G}} d\omega
\end{align}
where $\textbf{c}_\text{out} = \textbf{S}(\Omega_0) \textbf{c}_\text{in}$ with $\textbf{c}_\text{in}$ being a vector of $c^\text{in}_n$, $\textbf{c}_\text{out}$ being a vector of $c^\text{out}_\alpha$ and $\textbf{S}(\Omega)$ being a matrix given by:
\begin{subequations}\label{eq:smat}
\begin{align}
\big[\textbf{S}(\Omega)\big]_{\alpha, n} = -2\sqrt{2}\pi \sum_{m=1}^N \varepsilon_m^{\alpha^*} \xi_m^*(\Omega - \omega_\alpha)\times \nonumber \\ \big[\textbf{T}^{-1}(\Omega + i0^+)\textbf{X}(\Omega)\big]_{m, n}
\end{align}
where
\begin{align}
\big[\textbf{X}(\Omega)\big]_{m, n} = \int_{-\infty}^\infty \xi_m(\nu)\xi_m(\Omega - \nu) \xi_n^*(\nu)\xi_n^*(\Omega - \nu) d\nu.
\end{align}
\end{subequations}
The matrix $\textbf{S}(\Omega_0)$ maps the quantum state of an incoming two-photon wave-packet expressed on the basis of the scattering state overlaps ($\xi_n^*(\nu_1) \xi_n^*(\nu_2)$ for $n \in \{1, 2 \dots N\}$) to the trapped state expressed on the bound-state basis --- its inverse allows us to design the incident two-photon state (Eq.~\ref{eq:inc_two_ph_two_bs}) to excite a specific bound-state superposition. Furthermore, if the bound states are degenerate, i.e.~$\omega_\alpha = \omega_b$ for all $\alpha$, then $\ket{\psi_\text{out}}$, is separable into this bound superposition and a single-photon in the scattering state mode with spectrum $f_{\Delta, \Omega_0}(\omega + \omega_b)$. This allows heralding of a successful trapping process by detecting the scattered single-photon with a photon-number resolving detector.

As a concrete example of exciting bound state superpositions, we consider a time-delayed feedback system with two emitters (Fig.~\ref{fig:two_bs_ex}a). Assuming that both the emitters have the same resonance frequency $\omega_0$ and that $\omega_0 t_d = n\pi$ for some integer $n$, this system supports two bound states. Figure \ref{fig:two_bs_ex}b shows $\xi_1(\omega)$ and $\xi_2(\omega)$, the overlap of the scattering states with the two emitters. Figures~\ref{fig:two_bs_ex}c-e shows FDTD simulations of the response of this multi-emitter system to two-photon wave-packets that are designed using Eqs.~\ref{eq:inc_two_ph_two_bs} and \ref{eq:smat} to excite either of the two bound states individually (Fig.~\ref{fig:two_bs_ex}c-d) and an equal superposition of the two bound states (Fig.~\ref{fig:two_bs_ex}e).

In conclusion, using a scattering matrix formalism, we comprehensively studied the two-photon excitation of bound states in general non-Markovian wQED systems. We provided upper limits on the two-photon excitation probability of bound states, as well as the wave-packet that can achieve this upper limit. Furthermore, we also considered systems with multiple bound states and provided a formalism for constructing wave packets that can excite various superpositions of the bound states. The results in this paper not only further our understanding of bound state excitation in wQED systems, but also provide concrete quantum memory storage protocols using these systems.\\

\noindent\emph{Acknowledgements}: RT acknowledges support from the Kailath Graduate Fellowship and the Germany Research and Internship Programme (GRIP) award. DM acknowledges funding from ERC Advanced Grant QENOCOBA under the EU Horizon 2020 program (Grant Agreement No. 742102).  RT, SF and JV acknowledge funding from the Air Force Office of Scientific Research under AFOSR MURI programme (award no.~FA9550-17-1-0002). 

\bibliography{library.bib}

\begin{thebibliography}{28}%
\makeatletter
\providecommand \@ifxundefined [1]{%
 \@ifx{#1\undefined}
}%
\providecommand \@ifnum [1]{%
 \ifnum #1\expandafter \@firstoftwo
 \else \expandafter \@secondoftwo
 \fi
}%
\providecommand \@ifx [1]{%
 \ifx #1\expandafter \@firstoftwo
 \else \expandafter \@secondoftwo
 \fi
}%
\providecommand \natexlab [1]{#1}%
\providecommand \enquote  [1]{``#1''}%
\providecommand \bibnamefont  [1]{#1}%
\providecommand \bibfnamefont [1]{#1}%
\providecommand \citenamefont [1]{#1}%
\providecommand \href@noop [0]{\@secondoftwo}%
\providecommand \href [0]{\begingroup \@sanitize@url \@href}%
\providecommand \@href[1]{\@@startlink{#1}\@@href}%
\providecommand \@@href[1]{\endgroup#1\@@endlink}%
\providecommand \@sanitize@url [0]{\catcode `\\12\catcode `\$12\catcode
  `\&12\catcode `\#12\catcode `\^12\catcode `\_12\catcode `\%12\relax}%
\providecommand \@@startlink[1]{}%
\providecommand \@@endlink[0]{}%
\providecommand \url  [0]{\begingroup\@sanitize@url \@url }%
\providecommand \@url [1]{\endgroup\@href {#1}{\urlprefix }}%
\providecommand \urlprefix  [0]{URL }%
\providecommand \Eprint [0]{\href }%
\providecommand \doibase [0]{http://dx.doi.org/}%
\providecommand \selectlanguage [0]{\@gobble}%
\providecommand \bibinfo  [0]{\@secondoftwo}%
\providecommand \bibfield  [0]{\@secondoftwo}%
\providecommand \translation [1]{[#1]}%
\providecommand \BibitemOpen [0]{}%
\providecommand \bibitemStop [0]{}%
\providecommand \bibitemNoStop [0]{.\EOS\space}%
\providecommand \EOS [0]{\spacefactor3000\relax}%
\providecommand \BibitemShut  [1]{\csname bibitem#1\endcsname}%
\let\auto@bib@innerbib\@empty
\bibitem [{\citenamefont {Shen}\ and\ \citenamefont
  {Fan}(2007)}]{shen2007strongly}%
  \BibitemOpen
  \bibfield  {author} {\bibinfo {author} {\bibfnamefont {J.-T.}\ \bibnamefont
  {Shen}}\ and\ \bibinfo {author} {\bibfnamefont {S.}~\bibnamefont {Fan}},\
  }\href@noop {} {\bibfield  {journal} {\bibinfo  {journal} {Physical review
  letters}\ }\textbf {\bibinfo {volume} {98}},\ \bibinfo {pages} {153003}
  (\bibinfo {year} {2007})}\BibitemShut {NoStop}%
\bibitem [{\citenamefont {Shen}\ and\ \citenamefont
  {Fan}(2005)}]{shen2005coherent}%
  \BibitemOpen
  \bibfield  {author} {\bibinfo {author} {\bibfnamefont {J.-T.}\ \bibnamefont
  {Shen}}\ and\ \bibinfo {author} {\bibfnamefont {S.}~\bibnamefont {Fan}},\
  }\href@noop {} {\bibfield  {journal} {\bibinfo  {journal} {Physical review
  letters}\ }\textbf {\bibinfo {volume} {95}},\ \bibinfo {pages} {213001}
  (\bibinfo {year} {2005})}\BibitemShut {NoStop}%
\bibitem [{\citenamefont {Gonzalez-Ballestero}\ \emph
  {et~al.}(2014)\citenamefont {Gonzalez-Ballestero}, \citenamefont {Moreno},\
  and\ \citenamefont {Garcia-Vidal}}]{gonzalez2014generation}%
  \BibitemOpen
  \bibfield  {author} {\bibinfo {author} {\bibfnamefont {C.}~\bibnamefont
  {Gonzalez-Ballestero}}, \bibinfo {author} {\bibfnamefont {E.}~\bibnamefont
  {Moreno}}, \ and\ \bibinfo {author} {\bibfnamefont {F.}~\bibnamefont
  {Garcia-Vidal}},\ }\href@noop {} {\bibfield  {journal} {\bibinfo  {journal}
  {Physical Review A}\ }\textbf {\bibinfo {volume} {89}},\ \bibinfo {pages}
  {042328} (\bibinfo {year} {2014})}\BibitemShut {NoStop}%
\bibitem [{\citenamefont {Shi}\ \emph {et~al.}(2013)\citenamefont {Shi},
  \citenamefont {Fan} \emph {et~al.}}]{shi2013two}%
  \BibitemOpen
  \bibfield  {author} {\bibinfo {author} {\bibfnamefont {T.}~\bibnamefont
  {Shi}}, \bibinfo {author} {\bibfnamefont {S.}~\bibnamefont {Fan}},  \emph
  {et~al.},\ }\href@noop {} {\bibfield  {journal} {\bibinfo  {journal}
  {Physical Review A}\ }\textbf {\bibinfo {volume} {87}},\ \bibinfo {pages}
  {063818} (\bibinfo {year} {2013})}\BibitemShut {NoStop}%
\bibitem [{\citenamefont {Rephaeli}\ and\ \citenamefont
  {Fan}(2012)}]{rephaeli2012few}%
  \BibitemOpen
  \bibfield  {author} {\bibinfo {author} {\bibfnamefont {E.}~\bibnamefont
  {Rephaeli}}\ and\ \bibinfo {author} {\bibfnamefont {S.}~\bibnamefont {Fan}},\
  }\href@noop {} {\bibfield  {journal} {\bibinfo  {journal} {IEEE Journal of
  Selected Topics in Quantum Electronics}\ }\textbf {\bibinfo {volume} {18}},\
  \bibinfo {pages} {1754} (\bibinfo {year} {2012})}\BibitemShut {NoStop}%
\bibitem [{\citenamefont {Xu}\ \emph {et~al.}(2013)\citenamefont {Xu},
  \citenamefont {Rephaeli},\ and\ \citenamefont {Fan}}]{xu2013analytic}%
  \BibitemOpen
  \bibfield  {author} {\bibinfo {author} {\bibfnamefont {S.}~\bibnamefont
  {Xu}}, \bibinfo {author} {\bibfnamefont {E.}~\bibnamefont {Rephaeli}}, \ and\
  \bibinfo {author} {\bibfnamefont {S.}~\bibnamefont {Fan}},\ }\href@noop {}
  {\bibfield  {journal} {\bibinfo  {journal} {Physical review letters}\
  }\textbf {\bibinfo {volume} {111}},\ \bibinfo {pages} {223602} (\bibinfo
  {year} {2013})}\BibitemShut {NoStop}%
\bibitem [{\citenamefont {Xu}\ and\ \citenamefont {Fan}(2015)}]{xu2015input}%
  \BibitemOpen
  \bibfield  {author} {\bibinfo {author} {\bibfnamefont {S.}~\bibnamefont
  {Xu}}\ and\ \bibinfo {author} {\bibfnamefont {S.}~\bibnamefont {Fan}},\
  }\href@noop {} {\bibfield  {journal} {\bibinfo  {journal} {Physical Review
  A}\ }\textbf {\bibinfo {volume} {91}},\ \bibinfo {pages} {043845} (\bibinfo
  {year} {2015})}\BibitemShut {NoStop}%
\bibitem [{\citenamefont {Xu}\ and\ \citenamefont {Fan}(2017)}]{xu2017input}%
  \BibitemOpen
  \bibfield  {author} {\bibinfo {author} {\bibfnamefont {S.}~\bibnamefont
  {Xu}}\ and\ \bibinfo {author} {\bibfnamefont {S.}~\bibnamefont {Fan}},\ }in\
  \href@noop {} {\emph {\bibinfo {booktitle} {Quantum Plasmonics}}}\ (\bibinfo
  {publisher} {Springer},\ \bibinfo {year} {2017})\ pp.\ \bibinfo {pages}
  {1--23}\BibitemShut {NoStop}%
\bibitem [{\citenamefont {Trivedi}\ \emph {et~al.}(2018)\citenamefont
  {Trivedi}, \citenamefont {Fischer}, \citenamefont {Xu}, \citenamefont {Fan},\
  and\ \citenamefont {Vuckovic}}]{trivedi2018few}%
  \BibitemOpen
  \bibfield  {author} {\bibinfo {author} {\bibfnamefont {R.}~\bibnamefont
  {Trivedi}}, \bibinfo {author} {\bibfnamefont {K.}~\bibnamefont {Fischer}},
  \bibinfo {author} {\bibfnamefont {S.}~\bibnamefont {Xu}}, \bibinfo {author}
  {\bibfnamefont {S.}~\bibnamefont {Fan}}, \ and\ \bibinfo {author}
  {\bibfnamefont {J.}~\bibnamefont {Vuckovic}},\ }\href@noop {} {\bibfield
  {journal} {\bibinfo  {journal} {Physical Review B}\ }\textbf {\bibinfo
  {volume} {98}},\ \bibinfo {pages} {144112} (\bibinfo {year}
  {2018})}\BibitemShut {NoStop}%
\bibitem [{\citenamefont {Caneva}\ \emph {et~al.}(2015)\citenamefont {Caneva},
  \citenamefont {Manzoni}, \citenamefont {Shi}, \citenamefont {Douglas},
  \citenamefont {Cirac},\ and\ \citenamefont {Chang}}]{caneva2015quantum}%
  \BibitemOpen
  \bibfield  {author} {\bibinfo {author} {\bibfnamefont {T.}~\bibnamefont
  {Caneva}}, \bibinfo {author} {\bibfnamefont {M.~T.}\ \bibnamefont {Manzoni}},
  \bibinfo {author} {\bibfnamefont {T.}~\bibnamefont {Shi}}, \bibinfo {author}
  {\bibfnamefont {J.~S.}\ \bibnamefont {Douglas}}, \bibinfo {author}
  {\bibfnamefont {J.~I.}\ \bibnamefont {Cirac}}, \ and\ \bibinfo {author}
  {\bibfnamefont {D.~E.}\ \bibnamefont {Chang}},\ }\href@noop {} {\bibfield
  {journal} {\bibinfo  {journal} {New Journal of Physics}\ }\textbf {\bibinfo
  {volume} {17}},\ \bibinfo {pages} {113001} (\bibinfo {year}
  {2015})}\BibitemShut {NoStop}%
\bibitem [{\citenamefont {Dinc}\ \emph {et~al.}(2019)\citenamefont {Dinc},
  \citenamefont {Ercan},\ and\ \citenamefont {Bra{\'n}czyk}}]{dinc2019exact}%
  \BibitemOpen
  \bibfield  {author} {\bibinfo {author} {\bibfnamefont {F.}~\bibnamefont
  {Dinc}}, \bibinfo {author} {\bibfnamefont {{\.I}.}~\bibnamefont {Ercan}}, \
  and\ \bibinfo {author} {\bibfnamefont {A.~M.}\ \bibnamefont {Bra{\'n}czyk}},\
  }\href@noop {} {\bibfield  {journal} {\bibinfo  {journal} {Quantum}\ }\textbf
  {\bibinfo {volume} {3}},\ \bibinfo {pages} {213} (\bibinfo {year}
  {2019})}\BibitemShut {NoStop}%
\bibitem [{\citenamefont {Fang}\ \emph {et~al.}(2015)\citenamefont {Fang},
  \citenamefont {Baranger} \emph {et~al.}}]{fang2015waveguide}%
  \BibitemOpen
  \bibfield  {author} {\bibinfo {author} {\bibfnamefont {Y.-L.~L.}\
  \bibnamefont {Fang}}, \bibinfo {author} {\bibfnamefont {H.~U.}\ \bibnamefont
  {Baranger}},  \emph {et~al.},\ }\href@noop {} {\bibfield  {journal} {\bibinfo
   {journal} {Physical Review A}\ }\textbf {\bibinfo {volume} {91}},\ \bibinfo
  {pages} {053845} (\bibinfo {year} {2015})}\BibitemShut {NoStop}%
\bibitem [{\citenamefont {Shi}\ \emph {et~al.}(2015)\citenamefont {Shi},
  \citenamefont {Chang},\ and\ \citenamefont {Cirac}}]{shi2015multiphoton}%
  \BibitemOpen
  \bibfield  {author} {\bibinfo {author} {\bibfnamefont {T.}~\bibnamefont
  {Shi}}, \bibinfo {author} {\bibfnamefont {D.~E.}\ \bibnamefont {Chang}}, \
  and\ \bibinfo {author} {\bibfnamefont {J.~I.}\ \bibnamefont {Cirac}},\
  }\href@noop {} {\bibfield  {journal} {\bibinfo  {journal} {Physical Review
  A}\ }\textbf {\bibinfo {volume} {92}},\ \bibinfo {pages} {053834} (\bibinfo
  {year} {2015})}\BibitemShut {NoStop}%
\bibitem [{\citenamefont {Pichler}\ and\ \citenamefont
  {Zoller}(2016)}]{pichler2016photonic}%
  \BibitemOpen
  \bibfield  {author} {\bibinfo {author} {\bibfnamefont {H.}~\bibnamefont
  {Pichler}}\ and\ \bibinfo {author} {\bibfnamefont {P.}~\bibnamefont
  {Zoller}},\ }\href@noop {} {\bibfield  {journal} {\bibinfo  {journal}
  {Physical review letters}\ }\textbf {\bibinfo {volume} {116}},\ \bibinfo
  {pages} {093601} (\bibinfo {year} {2016})}\BibitemShut {NoStop}%
\bibitem [{\citenamefont {Gonz{\'a}lez-Tudela}\ and\ \citenamefont
  {Cirac}(2017)}]{gonzalez2017quantum}%
  \BibitemOpen
  \bibfield  {author} {\bibinfo {author} {\bibfnamefont {A.}~\bibnamefont
  {Gonz{\'a}lez-Tudela}}\ and\ \bibinfo {author} {\bibfnamefont {J.~I.}\
  \bibnamefont {Cirac}},\ }\href@noop {} {\bibfield  {journal} {\bibinfo
  {journal} {Physical Review Letters}\ }\textbf {\bibinfo {volume} {119}},\
  \bibinfo {pages} {143602} (\bibinfo {year} {2017})}\BibitemShut {NoStop}%
\bibitem [{\citenamefont {Calaj{\'o}}\ \emph {et~al.}(2016)\citenamefont
  {Calaj{\'o}}, \citenamefont {Ciccarello}, \citenamefont {Chang},\ and\
  \citenamefont {Rabl}}]{calajo2016atom}%
  \BibitemOpen
  \bibfield  {author} {\bibinfo {author} {\bibfnamefont {G.}~\bibnamefont
  {Calaj{\'o}}}, \bibinfo {author} {\bibfnamefont {F.}~\bibnamefont
  {Ciccarello}}, \bibinfo {author} {\bibfnamefont {D.}~\bibnamefont {Chang}}, \
  and\ \bibinfo {author} {\bibfnamefont {P.}~\bibnamefont {Rabl}},\ }\href@noop
  {} {\bibfield  {journal} {\bibinfo  {journal} {Physical Review A}\ }\textbf
  {\bibinfo {volume} {93}},\ \bibinfo {pages} {033833} (\bibinfo {year}
  {2016})}\BibitemShut {NoStop}%
\bibitem [{\citenamefont {Calaj{\'o}}\ \emph {et~al.}(2019)\citenamefont
  {Calaj{\'o}}, \citenamefont {Fang}, \citenamefont {Baranger}, \citenamefont
  {Ciccarello} \emph {et~al.}}]{calajo2019exciting}%
  \BibitemOpen
  \bibfield  {author} {\bibinfo {author} {\bibfnamefont {G.}~\bibnamefont
  {Calaj{\'o}}}, \bibinfo {author} {\bibfnamefont {Y.-L.~L.}\ \bibnamefont
  {Fang}}, \bibinfo {author} {\bibfnamefont {H.~U.}\ \bibnamefont {Baranger}},
  \bibinfo {author} {\bibfnamefont {F.}~\bibnamefont {Ciccarello}},  \emph
  {et~al.},\ }\href@noop {} {\bibfield  {journal} {\bibinfo  {journal}
  {Physical review letters}\ }\textbf {\bibinfo {volume} {122}},\ \bibinfo
  {pages} {073601} (\bibinfo {year} {2019})}\BibitemShut {NoStop}%
\bibitem [{\citenamefont {Carmele}\ \emph {et~al.}(2020)\citenamefont
  {Carmele}, \citenamefont {Nemet}, \citenamefont {Canela},\ and\ \citenamefont
  {Parkins}}]{carmele2020pronounced}%
  \BibitemOpen
  \bibfield  {author} {\bibinfo {author} {\bibfnamefont {A.}~\bibnamefont
  {Carmele}}, \bibinfo {author} {\bibfnamefont {N.}~\bibnamefont {Nemet}},
  \bibinfo {author} {\bibfnamefont {V.}~\bibnamefont {Canela}}, \ and\ \bibinfo
  {author} {\bibfnamefont {S.}~\bibnamefont {Parkins}},\ }\href@noop {}
  {\bibfield  {journal} {\bibinfo  {journal} {Physical Review Research}\
  }\textbf {\bibinfo {volume} {2}},\ \bibinfo {pages} {013238} (\bibinfo {year}
  {2020})}\BibitemShut {NoStop}%
\bibitem [{\citenamefont {Sinha}\ \emph
  {et~al.}(2020{\natexlab{a}})\citenamefont {Sinha}, \citenamefont {Meystre},
  \citenamefont {Goldschmidt}, \citenamefont {Fatemi}, \citenamefont
  {Rolston},\ and\ \citenamefont {Solano}}]{sinha2020non}%
  \BibitemOpen
  \bibfield  {author} {\bibinfo {author} {\bibfnamefont {K.}~\bibnamefont
  {Sinha}}, \bibinfo {author} {\bibfnamefont {P.}~\bibnamefont {Meystre}},
  \bibinfo {author} {\bibfnamefont {E.~A.}\ \bibnamefont {Goldschmidt}},
  \bibinfo {author} {\bibfnamefont {F.~K.}\ \bibnamefont {Fatemi}}, \bibinfo
  {author} {\bibfnamefont {S.~L.}\ \bibnamefont {Rolston}}, \ and\ \bibinfo
  {author} {\bibfnamefont {P.}~\bibnamefont {Solano}},\ }\href@noop {}
  {\bibfield  {journal} {\bibinfo  {journal} {Physical review letters}\
  }\textbf {\bibinfo {volume} {124}},\ \bibinfo {pages} {043603} (\bibinfo
  {year} {2020}{\natexlab{a}})}\BibitemShut {NoStop}%
\bibitem [{\citenamefont {Dinc}\ and\ \citenamefont
  {Bra{\'n}czyk}(2019)}]{dinc2019non}%
  \BibitemOpen
  \bibfield  {author} {\bibinfo {author} {\bibfnamefont {F.}~\bibnamefont
  {Dinc}}\ and\ \bibinfo {author} {\bibfnamefont {A.~M.}\ \bibnamefont
  {Bra{\'n}czyk}},\ }\href@noop {} {\bibfield  {journal} {\bibinfo  {journal}
  {Physical Review Research}\ }\textbf {\bibinfo {volume} {1}},\ \bibinfo
  {pages} {032042} (\bibinfo {year} {2019})}\BibitemShut {NoStop}%
\bibitem [{\citenamefont {Sinha}\ \emph
  {et~al.}(2020{\natexlab{b}})\citenamefont {Sinha}, \citenamefont
  {Gonz{\'a}lez-Tudela}, \citenamefont {Lu},\ and\ \citenamefont
  {Solano}}]{sinha2020collective}%
  \BibitemOpen
  \bibfield  {author} {\bibinfo {author} {\bibfnamefont {K.}~\bibnamefont
  {Sinha}}, \bibinfo {author} {\bibfnamefont {A.}~\bibnamefont
  {Gonz{\'a}lez-Tudela}}, \bibinfo {author} {\bibfnamefont {Y.}~\bibnamefont
  {Lu}}, \ and\ \bibinfo {author} {\bibfnamefont {P.}~\bibnamefont {Solano}},\
  }\href@noop {} {\bibfield  {journal} {\bibinfo  {journal} {arXiv preprint
  arXiv:2006.12569}\ } (\bibinfo {year} {2020}{\natexlab{b}})}\BibitemShut
  {NoStop}%
\bibitem [{\citenamefont {Kockum}\ \emph {et~al.}(2018)\citenamefont {Kockum},
  \citenamefont {Johansson},\ and\ \citenamefont
  {Nori}}]{kockum2018decoherence}%
  \BibitemOpen
  \bibfield  {author} {\bibinfo {author} {\bibfnamefont {A.~F.}\ \bibnamefont
  {Kockum}}, \bibinfo {author} {\bibfnamefont {G.}~\bibnamefont {Johansson}}, \
  and\ \bibinfo {author} {\bibfnamefont {F.}~\bibnamefont {Nori}},\ }\href@noop
  {} {\bibfield  {journal} {\bibinfo  {journal} {Physical review letters}\
  }\textbf {\bibinfo {volume} {120}},\ \bibinfo {pages} {140404} (\bibinfo
  {year} {2018})}\BibitemShut {NoStop}%
\bibitem [{\citenamefont {Mirza}\ and\ \citenamefont
  {Schotland}(2016)}]{mirza2016multiqubit}%
  \BibitemOpen
  \bibfield  {author} {\bibinfo {author} {\bibfnamefont {I.~M.}\ \bibnamefont
  {Mirza}}\ and\ \bibinfo {author} {\bibfnamefont {J.~C.}\ \bibnamefont
  {Schotland}},\ }\href@noop {} {\bibfield  {journal} {\bibinfo  {journal}
  {Physical Review A}\ }\textbf {\bibinfo {volume} {94}},\ \bibinfo {pages}
  {012302} (\bibinfo {year} {2016})}\BibitemShut {NoStop}%
\bibitem [{\citenamefont {Zheng}\ and\ \citenamefont
  {Baranger}(2013)}]{zheng2013persistent}%
  \BibitemOpen
  \bibfield  {author} {\bibinfo {author} {\bibfnamefont {H.}~\bibnamefont
  {Zheng}}\ and\ \bibinfo {author} {\bibfnamefont {H.~U.}\ \bibnamefont
  {Baranger}},\ }\href@noop {} {\bibfield  {journal} {\bibinfo  {journal}
  {Physical review letters}\ }\textbf {\bibinfo {volume} {110}},\ \bibinfo
  {pages} {113601} (\bibinfo {year} {2013})}\BibitemShut {NoStop}%
\bibitem [{\citenamefont {Ramos}\ \emph {et~al.}(2016)\citenamefont {Ramos},
  \citenamefont {Vermersch}, \citenamefont {Hauke}, \citenamefont {Pichler},\
  and\ \citenamefont {Zoller}}]{ramos2016non}%
  \BibitemOpen
  \bibfield  {author} {\bibinfo {author} {\bibfnamefont {T.}~\bibnamefont
  {Ramos}}, \bibinfo {author} {\bibfnamefont {B.}~\bibnamefont {Vermersch}},
  \bibinfo {author} {\bibfnamefont {P.}~\bibnamefont {Hauke}}, \bibinfo
  {author} {\bibfnamefont {H.}~\bibnamefont {Pichler}}, \ and\ \bibinfo
  {author} {\bibfnamefont {P.}~\bibnamefont {Zoller}},\ }\href@noop {}
  {\bibfield  {journal} {\bibinfo  {journal} {Physical Review A}\ }\textbf
  {\bibinfo {volume} {93}},\ \bibinfo {pages} {062104} (\bibinfo {year}
  {2016})}\BibitemShut {NoStop}%
\bibitem [{\citenamefont {Pichler}\ \emph {et~al.}(2017)\citenamefont
  {Pichler}, \citenamefont {Choi}, \citenamefont {Zoller},\ and\ \citenamefont
  {Lukin}}]{pichler2017universal}%
  \BibitemOpen
  \bibfield  {author} {\bibinfo {author} {\bibfnamefont {H.}~\bibnamefont
  {Pichler}}, \bibinfo {author} {\bibfnamefont {S.}~\bibnamefont {Choi}},
  \bibinfo {author} {\bibfnamefont {P.}~\bibnamefont {Zoller}}, \ and\ \bibinfo
  {author} {\bibfnamefont {M.~D.}\ \bibnamefont {Lukin}},\ }\href@noop {}
  {\bibfield  {journal} {\bibinfo  {journal} {Proceedings of the National
  Academy of Sciences}\ ,\ \bibinfo {pages} {201711003}} (\bibinfo {year}
  {2017})}\BibitemShut {NoStop}%
\bibitem [{sup()}]{supp}%
  \BibitemOpen
  \href@noop {} {\ }\bibinfo {note} {See Supplemental Material at \emph{URL
  will be inserted by publisher} for detailed discussion of diagonalization of
  $H_0$, calculation of the two-photon scattering matrix $S_\alpha(\omega;
  \nu_1, \nu_2)$, calculation of overlap of the emitters with scattering states
  and bound states for multi-emitter systems with time-delayed feedback and
  formulation of FDTD for these systems.}\BibitemShut {Stop}%
\bibitem [{\citenamefont {Fang}(2019)}]{fang2019fdtd}%
  \BibitemOpen
  \bibfield  {author} {\bibinfo {author} {\bibfnamefont {Y.-L.~L.}\
  \bibnamefont {Fang}},\ }\href@noop {} {\bibfield  {journal} {\bibinfo
  {journal} {Computer Physics Communications}\ }\textbf {\bibinfo {volume}
  {235}},\ \bibinfo {pages} {422} (\bibinfo {year} {2019})}\BibitemShut
  {NoStop}%
\end{thebibliography}%
\end{document}


\title{Supplement to "Optimal two-photon excitation of bound states in non-Markovian waveguide QED"}
\author{Rahul Trivedi$^{1, 2}$}
\email{rtrivedi@stanford.edu}
\author{Daniel Malz$^{3, 4}$}
\author{Shanhui Fan$^{1,2}$}
\author{Jelena Vu\v{c}kovi\'c$^{1,2}$}
\affiliation{{$^1$E. L. Ginzton Laboratory, Stanford University, Stanford, CA 94305, USA. \\
	         $^2$Department of Electrical Engineering, Stanford, CA 94305, USA.\\
	         $^3$Max-Planck-Institute for Quantum Optics, Hans-Kopfermann-Str.~1, 85748 Garching, Germany.\\
	         $^4$Munich Center for Quantum Science and Technology (MCQST), Schellingstra{\ss}e 4, 80799 M{\"u}nchen, Germany.}}

\date{\today}
\maketitle

\section{Diagonalizing the quadratic part of the non-Markovian waveguide QED Hamiltonian}
Here, we provide a general recipe for calculating bound states and scattering states of non-Markovian waveguide QED systems. As described in the main-text, we consider a multi-emitter system with Hamiltonian $H_0$ given by:
\begin{align}
H_0 = \sum_{n=1}^N \omega_n \sigma_n^\dagger \sigma_n + \int_{-\infty}^\infty \omega s_\omega^\dagger s_\omega d\omega + \int_{-\infty}^\infty \sum_{n=1}^N \big(V_n(\omega) s_\omega \sigma_n^\dagger + V_n^*(\omega) \sigma_n s_\omega^\dagger) \frac{d\omega}{\sqrt{2\pi}}.
\end{align}
For the purpose of our analysis, it is more convenient to rewrite this Hamiltonian in terms of the position domain annihilation operator $s_x$ defined by
\begin{align}\label{eq:pos_dom_op}
s_x = \int_{-\infty}^\infty s_\omega \exp(i\omega x)\frac{d\omega}{\sqrt{2\pi}},
\end{align}
in terms of which $H_0$ can be expressed as:
\begin{align}
H_0 = \sum_{n=1}^N \omega_n \sigma_n^\dagger \sigma_n - i\int_{-\infty}^\infty s_x^\dagger \frac{\partial s_x}{\partial x} dx + \int_{-\infty}^\infty \sum_{n=1}^N \big(\mathcal{V}_n(x)s_x \sigma_n^\dagger + \mathcal{V}_n^*(x) s_x^\dagger \sigma_n\big) dx,
\end{align}
where
\begin{align}
\mathcal{V}_n(x) = \int_{-\infty}^\infty V_n(\omega) \exp(-i\omega x)\frac{d\omega}{2\pi}
\end{align}
\noindent \emph{Scattering states}: We first consider the calculation of the scattering states for this system at frequency $\omega$ described by annihilation operators $\psi_\omega$. We assume the following ansatz for $\psi_\omega$:
\begin{align}\label{eq:scat_state_ansatz}
\psi_\omega = \sum_{n=1}^N \beta_n(\omega) \sigma_n + \int_{-\infty}^\infty \Psi_\omega(x) s_x dx,
\end{align}
where $\beta_n(\omega)$ and $\Psi_\omega(x)$ are to be determined. By definition, $\psi_\omega$ should describe an eigen-mode of $H_0$ oscillating at frequency $\omega$ and therefore, $[\psi_\omega, H_0] = \omega \psi_\omega$. This yields
\begin{subequations}
\begin{align}
&i\frac{\partial \Psi_\omega(x)}{\partial x} + \sum_{n=1}^N \mathcal{V}_n(x) \beta_n(\omega) = \omega \Psi_\omega(x)\label{eq:scat_state_eq_1} \\
&\omega_n \beta_n(\omega) + \int_{-\infty}^\infty \mathcal{V}_n^*(x) \Psi_\omega(x) dx = \omega \beta_n(\omega) \ \text{for} \ n \in \{1, 2 \dots N\}\label{eq:scat_state_eq_2}
\end{align}
\end{subequations}
We choose $\Psi_\omega(x) = e^{-i\omega x}/\sqrt{2\pi}$ as $x\to -\infty$. This is equivalent to choosing the scattering state to be a plane wave at positions in the waveguide before it encounters the emitters. With this boundary condition, Eq.~\ref{eq:scat_state_eq_1} can be formally integrated to obtain
\begin{align}\label{eq:psi_in_terms_beta}
\Psi_\omega(x) = \frac{e^{-i\omega x}}{\sqrt{2\pi}} + i \sum_{n=1}^N \beta_n(\omega) \int_{-\infty}^\infty \mathcal{V}_n(y)e^{-i\omega(x - y)} \Theta(y \leq x) dy.
\end{align}
Substituting this into Eq.~\ref{eq:scat_state_eq_2}, we obtain a system of linear equations for $\beta_n(\omega)$:
\begin{subequations}
\begin{align}\label{eq:scat_state_mat}
\textbf{M}(\omega) \textbf{b}(\omega) = \textbf{f}(\omega)
\end{align}
where $\textbf{b}(\omega) = [\beta_1(\omega), \beta_2(\omega) \dots \beta_N(\omega)]^\text{T}$ and
\begin{align}
&[\textbf{M}(\omega)]_{m, n} = (\omega_n - \omega) \delta_{m, n} + i\int_{-\infty}^\infty \int_{-\infty}^\infty \mathcal{V}_m^*(x) \mathcal{V}_n(y) e^{-i\omega(x - y)}\Theta(y \leq x)dxdy \\
&[\textbf{s}(\omega)]_n = -\int_{-\infty}^\infty \mathcal{V}_n^*(x) e^{-i\omega x} \frac{dx}{\sqrt{2\pi}} = -\frac{V_n^*(\omega)}{\sqrt{2\pi}}
\end{align}
\end{subequations}
Eq.~\ref{eq:scat_state_mat} can be solved to obtain $\beta_1(\omega), \beta_2(\omega) \dots \beta_N(\omega)$ which can then be substituted into Eq.~\ref{eq:psi_in_terms_beta} to obtain $\Psi_\omega(x)$. This completes the calculation of the scattering state.

It is also useful to consider form of $\Psi_\omega(x)$ as $x \to \infty$. It follows from Eq.~\ref{eq:psi_in_terms_beta} that
\begin{align}\label{eq:psi_asymp}
\Psi_\omega(x) \to \frac{\tau(\omega)}{\sqrt{2\pi}} e^{-i\omega x} \ \text{as }x\to \infty,
\end{align}
where $\tau(\omega) = 1 + 2\pi i \sum_{n=1}^N \beta_n(\omega) V_n(\omega)$ can be interpreted as the transmission of a plane-wave incident on the waveguide. Since we are considering systems with only one waveguide, it is expected that $|\tau(\omega)| = 1$. To see this rigorously, we note that it follows from Eq.~\ref{eq:scat_state_eq_1} that
\begin{align}
\frac{\partial}{\partial x}|\Psi_\omega(x)|^2 = i\sum_{n=1}^N \bigg(\Psi_\omega^*(x) \mathcal{V}_n(x) \beta_n(\omega) - \Psi_\omega(x) \mathcal{V}_n^*(x) \beta_n^*(\omega)\bigg).
\end{align}
Integrating this from $-\infty$ to $\infty$ and using Eq.~\ref{eq:scat_state_eq_2}, we obtain
\begin{align}
\frac{1}{2\pi}\big(|\tau(\omega)|^2 - 1 \big) &= i\sum_{n = 1}^N \bigg(\beta_n(\omega) \int_{-\infty}^\infty \Psi_\omega^*(x) \mathcal{V}_n(x) dx - \beta_n^*(\omega) \int_{-\infty}^\infty \Psi_\omega(x) \mathcal{V}_n^*(x)\bigg) = 0,
\end{align}
which immediately implies that $|\tau(\omega)| = 1$.
 \\

\noindent \emph{Bound states}: Next we consider the calculation of the bound states and their frequencies. Following the notation introduced in the main text, we denote the annihilation operator for the bound state at frequency $\omega_\alpha$ by $\phi_\alpha$. We assume the following ansatz for the bound state:
\begin{align}\label{eq:bound_state_ansatz}
\phi_\alpha = \sum_{n=1}^N v_n^\alpha \sigma_n + \int_{-\infty}^\infty \Phi_\alpha(x) s_x dx
\end{align}
where $v_n^\alpha$ and $\Phi_\alpha(x)$ are to be determined. By definition, $\phi_\alpha$ described an eigen-mode of $H_0$ oscillating at frequency $\omega_\alpha$ and therefore $[\phi_\alpha, H_0] = \omega_\alpha \phi_\alpha$. This yields
\begin{subequations}
\begin{align}
&i\frac{\partial \Phi_\alpha(x)}{\partial x} + \sum_{n=1}^N \mathcal{V}_n(x) v_n^\alpha = \omega_\alpha \phi_\alpha(x) \ \text{and} \label{eq:bs_eq_1}\\
&\omega_n v_n^\alpha + \int_{-\infty}^\infty \mathcal{V}_n^*(x) \Phi_\alpha(x)dx = \omega_\alpha v_n^\alpha \ \text{for} \ n \in \{1, 2 \dots N\}\label{eq:bs_eq_2}
\end{align}
\end{subequations}
Furthermore, since the bound state is by definition normalizable, $[\phi_\alpha, \phi_\alpha^\dagger] = 1$. Equivalently,
\begin{align}\label{eq:bs_norm}
\sum_{n=1}^N |v_n^\alpha|^2 + \int_{-\infty}^\infty |\Phi_\alpha(x)|^2dx = 1.
\end{align}
Therefore $\Phi_\alpha(x)$ has a bounded $L^2$ norm and therefore $\Phi_\alpha(x) \to 0$ as $|x|\to \infty$. Eq.~\ref{eq:bs_eq_1} can then be integrated to obtain
\begin{align}\label{eq:phi_t_v}
\Phi_\alpha(x) = i\sum_{n=1}^N v_n^\alpha \int_{-\infty}^\infty \mathcal{V}_n(y) e^{-i\omega(x - y)} \Theta(y \leq x) dy.
\end{align}
Substituting this into Eq.~\ref{eq:bs_eq_2}, we obtain the following homogeneous system of equations for $v_n^\alpha$:
\begin{align}\label{eq:bs_mat_eq}
\textbf{M}(\omega_\alpha) \textbf{v}_\alpha = 0
\end{align}
where $\textbf{v}_\alpha = [v_1^\alpha, v_2^\alpha \dots v_N^\alpha]^\text{T}$ and $\textbf{M}(\omega)$ is defined in Eq.~\ref{eq:scat_state_mat}. Furthermore, since $\Phi_\alpha(x) \to 0$ as $x \to \infty$, it follows from Eq.~\ref{eq:phi_t_v} that
\begin{align}\label{eq:bs_consistency}
\sum_{n=1}^N v^\alpha_n V_n(\omega_\alpha) = 0.
\end{align}
Thus, to determine bound state frequencies, Eq.~\ref{eq:bs_mat_eq} implies that we must first solve $\text{det}[\textbf{M}(\omega_\alpha)] = 0$, followed by calculation of $v_n^\alpha$ again using Eq.~\ref{eq:bs_mat_eq}. Finally, the resulting solution for $\omega_\alpha$ and $v_n^\alpha$ needs to satisfy Eq.~\ref{eq:bs_consistency} to correspond to a normalizable bound state. It can also be noted that the overlap of the bound state with the waveguide mode, $\Phi_\alpha(x)$, can be computed using Eq.~\ref{eq:phi_t_v} and the normalization of the coefficients $v_n^\alpha$ can be fixed by using Eq.~\ref{eq:bs_norm}.\\

\noindent \emph{Commutation relations}: Since the bound states and scattering states are eigen-modes of the quadratic Hamiltonian $H_0$, they are expected to satisfy the commutations:
\begin{align}
[\phi_\alpha, \phi_\beta^\dagger] = \delta_{\alpha, \beta}, [\psi_\omega, \psi_\nu^\dagger] = \delta(\omega - \nu) \text{ and } [\phi_\alpha, \psi_\nu^\dagger] = 0.
\end{align}
Below, we show that the previously calculated bound states and scattering states do indeed satisfy these commutators.
\begin{enumerate}
\item \emph{Commutator between two scattering states}: We consider scattering state modes at frequency $\omega$ and $\nu$. From Eq.~\ref{eq:scat_state_ansatz}, it follows that
\begin{align}
[\psi_\omega, \psi_\nu^\dagger] = \sum_{n=1}^N \beta_n(\omega) \beta_n^*(\nu) + \int_{-\infty}^\infty \Psi_\omega(x) \Psi_\nu^*(x) dx.
\end{align}
For evaluating this commutator, it is useful to rewrite Eq.~\ref{eq:psi_in_terms_beta} as
\begin{subequations}\label{eq:beta_to_psi}
\begin{align}\label{eq:psi_as_scat}
\Psi_\omega(x) = \frac{e^{-i\omega x}}{\sqrt{2\pi}} + \tilde{\Psi}_\omega(x),
\end{align}
where
\begin{align}\label{eq:psi_scat}
\tilde{\Psi}_\omega(x) =  i\lim_{\mu \to 0}\bigg[\sum_{n=1}^N \beta_n(\omega) \int_{-\infty}^\infty \mathcal{V}_n(y)e^{-(i\omega + \mu)(x - y)} \Theta(y \leq x)dy\bigg].
\end{align}
\end{subequations}
Using Eq.~\ref{eq:beta_to_psi}, we then obtain
\begin{align}
\int_{-\infty}^\infty & \Psi_\omega(x) \Psi_\nu^*(x)  dx = \delta(\omega - \nu) + \int_{-\infty}^\infty \tilde{\Psi}_\omega(x) \tilde{\Psi}_\nu^*(x) dx + \int_{-\infty}^\infty \tilde{\Psi}_\omega(x)e^{i\nu x}\frac{dx}{\sqrt{2\pi}} + \int_{-\infty}^\infty \tilde{\Psi}_\nu^*(x) e^{-i\omega x}\frac{dx}{\sqrt{2\pi}}.
\end{align}
The various integrals in the above equation can be evaluated using Eq.~\ref{eq:psi_scat}. It follows that
\begin{align}\label{eq:int_1}
\int_{-\infty}^\infty \tilde{\Psi}_\omega(x) \tilde{\Psi}^*_\nu(x) dx &= \lim_{\mu \to 0}\sum_{n, m = 1}^N \frac{\beta_n(\omega) \beta_m^*(\nu)}{2\mu + i(\omega - \nu)} \int_{-\infty}^\infty \int_{-\infty}^\infty \mathcal{V}_n(y_1) \mathcal{V}_m^*(y_2) e^{-i(\omega - \nu) \text{max}(y_1, y_2)} e^{i(\omega y_1 - \nu y_2)} dy_1 dy_2\nonumber \\
&=\lim_{\mu \to 0}\sum_{n, m = 1}^N \frac{\beta_n(\omega) \beta_m^*(\nu)}{\omega - \nu - 2i\mu}\bigg( \big[\textbf{M}(\nu)\big]_{n, m}^* - \big[\textbf{M}(\omega)\big]_{m, n} - (\omega - \nu)\delta_{n, m}\bigg)\nonumber\\
&=\lim_{\mu \to 0}\sum_{n, m = 1}^N \frac{V_n^*(\omega) \beta_n^*(\nu) - V_n(\nu) \beta_n(\omega)}{\omega - \nu - 2i\mu} - \sum_{n = 1}^N \beta_n(\omega) \beta_n^*(\nu),
\end{align}
wherein in the last step we have used the fact that the coefficients $\beta_1(\omega), \beta_2(\omega) \dots \beta_N(\omega)$ satisfy the system of equations in Eq.~\ref{eq:scat_state_mat}. Similarly, we also obtain
\begin{align}\label{eq:int_2}
\int_{-\infty}^\infty \tilde{\Psi}_\omega(x) e^{i\nu x} \frac{dx}{\sqrt{2\pi}} = \lim_{\mu \to 0} \sum_{n = 1}^N \frac{V_n(\nu) \beta_n(\omega)}{\omega - \nu - i\mu},
\end{align}
and
\begin{align}\label{eq:int_3}
\int_{-\infty}^\infty \tilde{\Psi}_\nu(x) e^{-i\omega x} \frac{dx}{\sqrt{2\pi}} = -\lim_{\mu \to 0} \sum_{n = 1}^N \frac{V_n^*(\omega)\beta_n^*(\nu)}{\omega - \nu - i\mu}.
\end{align}
From Eqs.~\ref{eq:int_1}, \ref{eq:int_2} and \ref{eq:int_3}, it immediately follows that
\begin{align}
\int_{-\infty}^\infty \Psi_\omega(x) \Psi_\nu^*(x) dx = \delta(\omega - \nu) - \sum_{n = 1}^N \beta_n(\omega) \beta_n^*(\nu).
\end{align}
This immediately implies that the scattering state mode annihilation operators $\psi_\omega$ satisfy the commutator $[\psi_\omega, \psi_\nu^\dagger] = \delta(\omega - \nu)$.

\item \emph{Commutator between two bound states}: We consider two bound-state modes with annihilation operators $\phi_\alpha$ and $\phi_\beta$. We assume that these correspond to bound states at different frequencies ($\omega_\alpha \neq \omega_\beta$), since for degenerate modes ($\omega_\alpha = \omega_\beta$) the bound-states can be chosen to satisfy $[\phi_\alpha, \phi_\beta^\dagger] = \delta_{\alpha, \beta}$. Furthermore, we note that if the bound-state modes are normalized as per Eq.~\ref{eq:bs_norm} then the commutator $[\phi_\alpha, \phi_\alpha^\dagger] = 1$ is also satisfied. From Eq.~\ref{eq:bound_state_ansatz}, it follows that
\begin{align}\label{eq:comm_as_int}
[\phi_\alpha, \phi_\beta^\dagger] = \sum_{n=1}^N v_n^\alpha v_n^{\beta*} + \int_{-\infty}^\infty \Phi_\alpha(x) \Phi_\beta^*(x) dx.
\end{align}
Multiplying Eq.~\ref{eq:bs_eq_1} by $\Phi_\beta^*(x)$  and integrating from $-\infty$ to $\infty$ we obtain
\begin{subequations}
\begin{align}\label{eq:inner_prod_1}
i\int_{-\infty}^\infty \Phi_\beta^*(x) \frac{\partial \Phi_\alpha(x)}{\partial x} dx  + \sum_{n=1}^N v_n^\alpha \int_{-\infty}^\infty \Phi_\beta^*(x) \mathcal{V}_n(x) dx = \omega_\alpha \int_{-\infty}^\infty \Phi_\beta^*(x) \Phi_\alpha(x) dx
\end{align}
Switching $\alpha$ and $\beta$ in Eq.~\ref{eq:inner_prod_1} and conjugating, we obtain
\begin{align}\label{eq:inner_prod_2}
-i\int_{-\infty}^\infty \Phi_\alpha(x) \frac{\partial \Phi_\beta^*(x)}{\partial x} dx  + \sum_{n=1}^N v_n^{\beta*} \int_{-\infty}^\infty \Phi_\alpha(x) \mathcal{V}_n^*(x) dx = \omega_\beta \int_{-\infty}^\infty \Phi_\beta^*(x) \Phi_\alpha(x) dx
\end{align}
\end{subequations}
Subtracting Eqs.~\ref{eq:inner_prod_1} and \ref{eq:inner_prod_2}, we obtain
\begin{align}\label{eq:int_v1}
\sum_{n=1}^N\bigg(v_n^\alpha \int_{-\infty}^\infty \Phi_\beta^*(x) \mathcal{V}_n(x) dx -v_n^{\beta*} \int_{-\infty}^\infty \Phi_\alpha(x) \mathcal{V}_n^*(x) dx  \bigg) = (\omega_\alpha - \omega_\beta) \int_{-\infty}^\infty \Phi_\beta^*(x) \Phi_\alpha(x) dx,
\end{align}
wherein we have used the fact that $\Phi_\alpha(x), \Phi_\beta(x) \to 0$ as $|x| \to \infty$ to set
\begin{align}
\int_{-\infty}^\infty \bigg(\Phi_\beta^*(x) \frac{\partial \Phi_\alpha(x)}{\partial x} + \Phi_\alpha(x) \frac{\partial \Phi_\beta^*(x)}{\partial x}\bigg)dx = \int_{-\infty}^\infty \frac{\partial (\Phi_\beta^*(x) \Phi_\alpha(x))}{\partial x} dx = 0.
\end{align}
Furthermore, using Eqs.~\ref{eq:bs_eq_2} and \ref{eq:int_v1}, under the assumption that $\omega_\alpha \neq \omega_\beta$ we obtain
\begin{align}
\int_{-\infty}^\infty \Phi_\beta^*(x) \Phi_\alpha(x) dx = -\sum_{n = 1}^N v_n^\alpha v_n^{\beta*}.
\end{align}
This together with Eq.~\ref{eq:comm_as_int} immediately implies that $[\phi_\alpha, \phi_\beta^\dagger] = 0$ if the two bound state modes under consideration have different frequencies. This completes the derivation of the commutator $[\phi_\alpha, \phi_\beta^\dagger] = \delta_{\alpha, \beta}$.

\item \emph{Commutator between a bound state and a scattering state}: We consider a bound state mode at frequency $\omega_\alpha$ with annihilation operators $\phi_\alpha$ and a scattering state mode with operator $\psi_\omega$ at frequency $\omega \neq \omega_\alpha$. Using Eqs.~\ref{eq:scat_state_ansatz} and \ref{eq:bound_state_ansatz}, it follows that
\begin{align}
[\phi_\alpha, \psi_\omega^\dagger] = \sum_{n = 1}^N v_n^\alpha \beta_n^*(\omega) + \int_{-\infty}^\infty \Phi_\alpha(x) \Psi_\omega^*(x) dx
\end{align}
Following manipulations similar to those done for the case of two bound state modes, it is easily shown that
\begin{align}
\int_{-\infty}^\infty \Phi_\alpha(x) \Psi_\omega^*(x) dx = -\sum_{n = 1}^N v_n^\alpha \beta_n^*(\omega),
\end{align}
which immediately implies that $[\phi_\alpha, \psi_\omega^\dagger] = 0$ if $\omega \neq \omega_\alpha$. Since we expect $[\phi_\alpha, \psi_\omega^\dagger]$ to be a continuous function of $\omega$, it follows that $[\phi_\alpha, \psi_\omega^\dagger] = 0$ for all $\omega$.
\end{enumerate}
\noindent \emph{Expressing the lowering operators in terms of $\psi_\omega$ and $\phi_\alpha$}: Finally, we seek a representation of the lowering operator $\sigma_n$ in terms of $\psi_\omega$ and $\phi_\alpha$:
\begin{align}\label{eq:bs_ss_to_olap}
\sigma_n = \sum_{\alpha=1}^{N_b}\varepsilon_n^\alpha \phi_\alpha + \int_{-\infty}^\infty \xi_n(\omega) \psi_\omega d\omega.
\end{align}
This ansatz immediately yields that $\varepsilon_n^\alpha = [\sigma_n, \phi_\alpha^\dagger]$ and $\xi_n(\omega) = [\sigma_n, \psi_\omega^\dagger]$. From Eqs.~\ref{eq:scat_state_ansatz} and \ref{eq:bound_state_ansatz}, we then immediately obtain $\varepsilon_n^\alpha = v_n^{\alpha*}$ and $\xi_n(\omega) = \beta_n^*(\omega)$.

\section{Calculation of two-photon scattering matrix element}
The scattering matrix element that we are interested in calculating for understanding the bound-state trapping process is given by
\begin{align}
S_\alpha(\omega; \nu_1, \nu_2) =\lim_{\substack{t_i\to -\infty \\ t_f \to \infty}} \bra{\text{G}}\phi_\alpha \psi_\omega U_I(t_f, t_i) \psi_{\nu_1}^\dagger \psi_{\nu_2}^\dagger \ket{\text{G}}.
\end{align}
Since $U_I(t_f, t_i) = e^{iH_0 t_f} e^{-iH(t_f - t_i)} e^{-iH_0 t_i}$. The expectation value in the above equation can be expressed in terms of Heisenberg picture operators with respect to the Hamiltonian $H_0$:
\begin{align}
\bra{\text{G}}\phi_\alpha \psi_\omega U_I(t_f, t_i) \psi_{\nu_1}^\dagger \psi_{\nu_2}^\dagger \ket{\text{G}} = \bra{\text{G}}\mathcal{T}\bigg[\phi_\alpha(t_f) \psi_\omega(t_f) \exp\bigg(-\frac{iU_0}{2}\int_{t_i}^{t_f} \sum_{n=1}^N \big(\sigma^\dagger(\tau)\big)^2 \sigma^2(\tau) d\tau\bigg)\psi_{\nu_1}(t_i)\psi_{\nu_2}(t_i)\bigg]\ket{\text{G}},
\end{align}
where $\mathcal{T}$ is the time-ordering operator. A dyson series expansion of the exponential then yields
\begin{align}\label{eq:series_exp}
S_\alpha(\omega; \nu_1, \nu_2) = \sum_{k=1}^\infty \frac{1}{k!} \bigg(-\frac{iU_0}{2}\bigg)^k \mathcal{G}_\alpha^k(\omega; \nu_1, \nu_2),
\end{align}
where:
\begin{align}\label{eq:ind_term}
&\mathcal{G}^k_\alpha(\omega; \nu_1, \nu_2) \nonumber \\&  =\lim_{\substack{t_i\to-\infty \\ t_f \to \infty}}e^{i\theta(t_f, t_i)}\int_{\tau_1, \tau_2 \dots \tau_k = t_i}^{t_f} \bra{\text{G}}\mathcal{T}\bigg[\phi_\alpha(t_f) \psi_\omega(t_f) \psi_{\nu_1}^\dagger(t_i)\psi_{\nu_2}^\dagger(t_i)\prod_{n=1}^{k} \sum_{m=1}^N \big(\sigma_{m}^\dagger(\tau_n)\big)^2 \sigma_m^2(\tau_n)\bigg]\ket{\text{G}} d\tau_1 d\tau_2 \dots d\tau_k
\end{align}
with $\theta(t_f, t_i) = (\omega_\alpha + \omega) t_f - (\nu_1 + \nu_2) t_i$. The integral in Eq.~\ref{eq:ind_term} can be manipulated further to make it more amenable to analytical evaluation:
\begin{align}\label{eq:simp_int}
&\int_{\tau_1, \tau_2 \dots \tau_k = t_i}^{t_f} \bra{\text{G}}\mathcal{T}\bigg[\phi_\alpha(t_f) \psi_\omega(t_f) \psi_{\nu_1}^\dagger(t_i)\psi_{\nu_2}^\dagger(t_i)\prod_{n=1}^{k} \sum_{m=1}^N \big(\sigma_{m}^\dagger(\tau_n)\big)^2 \sigma_m^2(\tau_n)\bigg]\ket{\text{G}} d\tau_1 d\tau_2 \dots d\tau_k\nonumber \\
&= \int_{\tau_1, \tau_2\dots \tau_k = t_i}^{t_f}\sum_{m_1, m_2 \dots m_k = 1}^N \bra{\text{G}} \mathcal{T}\bigg[\phi_\alpha(t_f) \psi_\omega(t_f) \psi_{\nu_1}^\dagger(t_i) \psi_{\nu_2}^\dagger(t_i)\prod_{n=1}^k \big(\sigma_{m_n}^\dagger(\tau_n)\big)^2 \sigma_{m_n}^2(\tau_n)\bigg]\ket{\text{G}}d\tau_1 d\tau_2 \dots d\tau_k \nonumber \\
&=k!\int_{\tau_1 > \tau_2 \dots > \tau_k = t_i}^{t_f}\sum_{m_1, m_2 \dots m_k = 1}^N\bra{\text{G}} \phi_\alpha(t_f) \psi_\omega(t_f) \bigg[\prod_{n=1}^k \big(\sigma_{m_n}^\dagger(\tau_n)\big)^2 \sigma_{m_n}^2(\tau_n)\bigg] \psi_{\nu_1}^\dagger(t_i) \psi_{\nu_2}^\dagger(t_i)\ket{\text{G}}d\tau_1 d\tau_2 \dots d\tau_k,
\end{align}
wherein in the last step we have made use of the fact that the integrand is invariant under a permutation  of the time-indices $\tau_1, \tau_2 \dots \tau_k$. To further evaluate this integral, we note that since the operators $\phi_\alpha$ and $\psi_\omega$ diagonalize the Hamiltonian $H_0$, it immediately follows that
\begin{align}\label{eq:hpic_bs_scat_s}
\phi_\alpha(t) = e^{iH_0t} \phi_\alpha e^{-iH_0 t} = \phi_\alpha e^{-i\omega_\alpha t} \text{ and } \psi_\omega(t) = e^{iH_0 t} \psi_\omega e^{-iH_0 t} = \psi_\omega e^{-i\omega t}
\end{align}
Furthermore, it follows from Eq.~\ref{eq:bs_ss_to_olap} that
\begin{align}\label{eq:hpic_tls_op}
\sigma_n(t) = e^{iH_0 t}\sigma_n e^{-iH_0 t}  = \sum_{\alpha=1}^{N_b}\varepsilon_n^\alpha \phi_\alpha e^{-i\omega_\alpha t} + \int_{-\infty}^\infty \xi_n(\omega) e^{-i\omega t} d\omega
\end{align}
Eqs.~\ref{eq:hpic_bs_scat_s} and \ref{eq:hpic_tls_op} then yield the following commutation relations:
\begin{subequations}\label{eq:linear_comm}
\begin{align}
&[\phi_\alpha(t), \sigma_n^\dagger(s)] = \varepsilon_n^{\alpha*}e^{-i\omega_\alpha(t-s)} \\
&[\psi_\omega(t), \sigma_n^\dagger(s)] = \xi_n(\omega) e^{-i\omega(t-s)} \\
&[\sigma_m(t), \sigma_n^\dagger(s)] = G_{m, n}(t-s) = \sum_{n=1}^{N_b}\varepsilon_m^\alpha \varepsilon_n^{\alpha*}e^{-i\omega_\alpha(t-s)} + \int_{-\infty}^\infty \xi_m(\omega) \xi_n^*(\omega) e^{-i\omega (t-s)} d\omega
\end{align}
\end{subequations}
With these commutation relations, the evaluation of the expectation in Eq.~\ref{eq:simp_int} is easily done in the usual way by moving all the annihilation operators to the right and all the creation operators to the left to obtain:
\begin{align}
\bra{\text{G}} &\phi_\alpha(t_f) \psi_\omega(t_f) \bigg[\prod_{n=1}^k \big(\sigma_{m_n}^\dagger(\tau_n)\big)^2 \sigma_{m_n}^2(\tau_n)\bigg] \psi_{\nu_1}^\dagger(t_i) \psi_{\nu_2}^\dagger(t_i)\ket{\text{G}} \nonumber \\
&=e^{-i \theta(t_f, t_i)} 2^{k + 1}\varepsilon_{m_1}^{\alpha*}\xi_{m_1}^*(\omega)\xi_{m_k}(\nu_1) \xi_{m_k}(\nu_2)e^{-i(\omega + \omega_\alpha)\tau_1}e^{i(\nu_1 + \nu_2) \tau_k} \bigg[\prod_{n=1}^{k-1}G_{m_n, m_{n+1}}^2(\tau_n - \tau_{n + 1})\bigg].
\end{align}
Using this result along with Eqs.~\ref{eq:simp_int} and \ref{eq:ind_term}, we obtain:
\begin{align}
\mathcal{G}_\alpha^k(\omega; \nu_1, \nu_2) = 2^{k+2}\pi \delta(\omega + \omega_\alpha - \nu_1 - \nu_2) \sum_{m, n = 1}^N\big(\varepsilon^\alpha_m\big)^* \xi^*_m(\omega) \big[\textbf{T}^{k-1}(\omega + \omega_\alpha + i0^+)\big]_{m, n}\xi_n(\nu_1) \xi_n(\nu_2),
\end{align}
where $\textbf{T}(\Omega + i0^+)$ is defined in Eqs.~6c and 6d of the main text. Finally, substituting this expression for $\mathcal{G}_\alpha^k(\omega; \nu_1, \nu_2)$ into the series expansion in Eq.~\ref{eq:series_exp} yields the result for the scattering matrix in Eq.~6a and 6b of the main text.

\section{Bound on trapping probability}
Here, we give a short derivation of the bound on the trapping probability of a bound state on excitation with an incident two-photon wave-packet. On excitation with a two-photon wave-packet $\psi_\text{in}(\nu_1, \nu_2)$, the probability of trapping the $\alpha^\text{th}$ bound state is given by (Eq.~8 of main text):
\begin{align}
P_\alpha[\psi_\text{in}] = \frac{1}{2}\int_{-\infty}^\infty d\Omega \bigg| \int_{-\infty}^\infty \Gamma_\alpha(\Omega - \omega_\alpha; \nu, \Omega - \nu)\psi_\text{in}(\nu, \Omega - \nu)d\nu \bigg|^2
\end{align}
From the Cauchy-schwarz inequality, it follows that:
\begin{align}
P_\alpha[\psi_\text{in}] \leq \frac{1}{2}\int_{-\infty}^\infty d\Omega \bigg(\int_{-\infty}^\infty  \big|\Gamma_\alpha(\Omega - \omega_\alpha; \nu, \Omega - \nu)\big|^2 d\nu\bigg) \bigg(\int_{-\infty}^\infty |\psi_\text{in}(\nu, \Omega - \nu)|^2 d\nu\bigg).
\end{align}
Furthermore, since
\begin{align}
\bigg(\int_{-\infty}^\infty  \big|\Gamma_\alpha(\Omega - \omega_\alpha;& \nu, \Omega - \nu)\big|^2 d\nu\bigg) \bigg(\int_{-\infty}^\infty |\psi_\text{in}(\nu, \Omega - \nu)|^2 d\nu\bigg) \leq \nonumber \\ &\bigg[\max_{\Omega \in \mathbb{R}}\bigg(\int_{-\infty}^\infty \big|\Gamma_\alpha(\Omega - \omega_\alpha; \nu, \Omega - \nu)\big|^2d\nu \bigg)\bigg]\int_{-\infty}^\infty |\psi_\text{in}(\nu, \Omega - \nu)|^2 d\nu,
\end{align}
it follows that
\begin{align}\label{eq:bound_inter}
P_\alpha[\psi] \leq \frac{1}{2}\max_{\Omega \in \mathbb{R}}\bigg(\int_{-\infty}^\infty \big|\Gamma_\alpha(\Omega - \omega_\alpha; \nu, \Omega - \nu)\big|^2d\nu \bigg) \int_{\Omega, \nu -\infty}^\infty  |\psi_\text{in}(\nu, \Omega - \nu)|^2 d\nu\ d\Omega.
\end{align}
Furthermore, since $\psi_\text{in}(\nu_1, \nu_2)$ is normalized to 1, it follows that:
\begin{align}
\int_{\nu_1, \nu_2 = -\infty}^\infty |\psi_\text{in}(\nu_1, \nu_2)|^2 d\nu_1 d\nu_2 = \int_{\nu, \Omega = -\infty}^\infty |\psi_\text{in}(\nu, \Omega - \nu)|^2 d\nu\ d\Omega = 1.
\end{align}
Substituting this into Eq.~\ref{eq:bound_inter}, we obtain the upper bound in Eq.~9 of the main text.
\section{Multi-emitter time-delayed feedback systems}
\subsection{Scattering states and bound states}
\begin{figure}[b]
\centering
\includegraphics[scale=0.4]{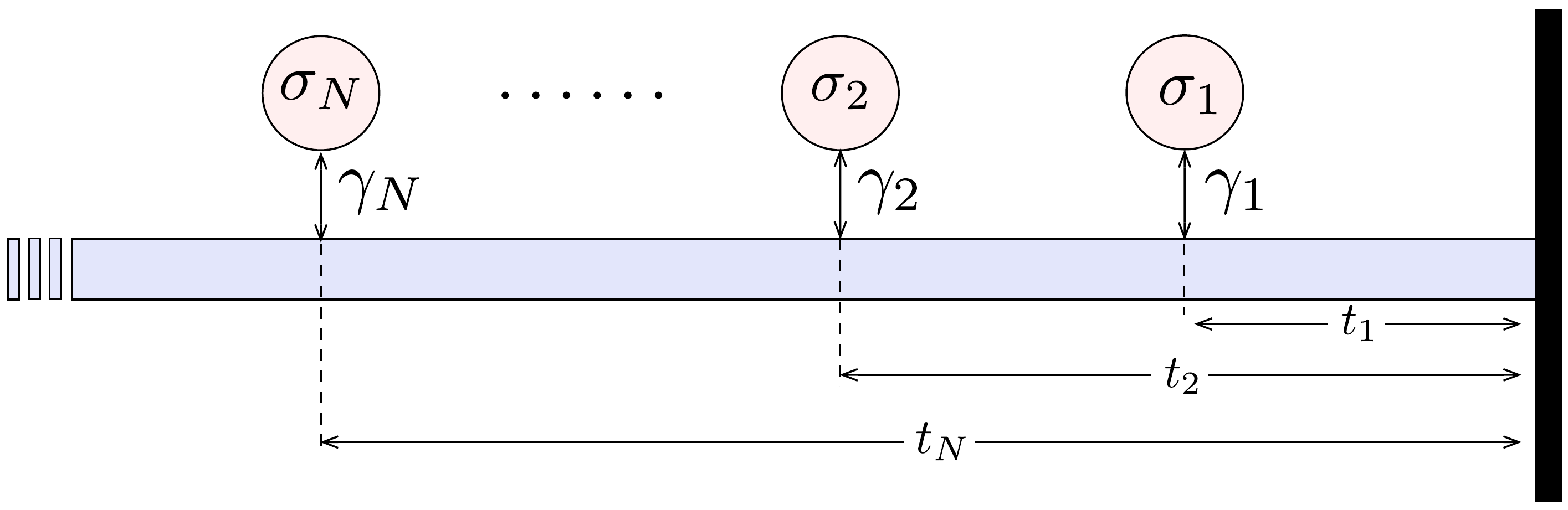}
\caption{Schematic of a multi-emitter waveguide-QED system with time-delayed feedback.}
\label{fig:supp_schematic}
\end{figure}
In this section, we consider the calculation of bound states and scattering states of multi-emitter time-delayed feedback systems. We consider system shown in Fig.~\ref{fig:supp_schematic} --- $N$ emitters with lowering operators $\sigma_1, \sigma_2 \dots \sigma_N$ are coupled to the forward and backward propagating modes of the waveguide with decay rates $\gamma_1, \gamma_2 \dots \gamma_N$. The waveguide mode is terminated with a perfect mirror which is at a distance $t_n$ from the $n^\text{th}$ emitter. The quadratic part of the Hamiltonian for this system, $H_0$, can be expressed as
\begin{align}
H_0 = \sum_{n=1}^N \omega_n \sigma_n^\dagger \sigma_n + \int_{-\infty}^\infty \omega s_\omega^\dagger s_\omega d\omega -2i \int_{-\infty}^\infty \sum_{n=1}^N \sqrt{\gamma_n}\sin(\omega t_n) \big(s_\omega \sigma_n^\dagger - s_\omega^\dagger \sigma_n) \frac{d\omega}{\sqrt{2\pi}},
\end{align}
where $\omega_n$ is the resonance frequency of the $n^\text{th}$ emitter. Alternatively, this hamiltonian can be expressed in terms of the position domain annihilation operator defined in Eq.~\ref{eq:pos_dom_op}:
\begin{align}
H_0 = \sum_{n=1}^N \omega_n \sigma_n^\dagger \sigma_n - i\int_{-\infty}^\infty s_x^\dagger \frac{\partial s_x}{\partial x} dx + \sum_{n=1}^N\sqrt{\gamma_n} \big[(s_{x = -t_n} - s_{x = t_n}) \sigma_n^\dagger + \text{h.c.}\big]
\end{align}
We first consider the calculation of the scattering states for this system. Assuming the ansatz in Eq.~\ref{eq:scat_state_ansatz} for the scattering state annihilation operator $\psi_\omega$ and using Eqs.~\ref{eq:scat_state_eq_1} and \ref{eq:scat_state_eq_2}, we obtain
\begin{subequations}
\begin{align}
&i\frac{\partial \Psi_\omega(x)}{\partial x} + \sum_{n=1}^N \sqrt{\gamma_n}\big(\delta(x + t_n) - \delta(x - t_n)\big) \beta_n(\omega) = \omega \Psi_\omega(x), \label{eq:scat_state_tdf_1}\\
&\omega_n \beta_n(\omega) + \sqrt{\gamma_n}\big(\Psi_\omega(-t_n) - \Psi_\omega(t_n)\big) = \omega \beta_n(\omega). \label{eq:scat_state_tdf_2}
\end{align}
\end{subequations}
With the boundary condition $\Psi_\omega(x) \to e^{-i\omega x}/\sqrt{2\pi}$ as $x \to -\infty$, the solution to Eq.~\ref{eq:scat_state_tdf_1} can be expressed as
\begin{align}
\Psi_\omega(x) = \frac{e^{-i\omega x}}{\sqrt{2\pi}}\begin{cases}
1 & \text{ for } x < -t_N, \\
C_{n}^- & \text{ for } -t_{n + 1} < x < -t_{n},\ n \in \{1, 2, \dots N - 1\}, \\
C_0  & \text{ for } -t_1 < x < t_1,\\
C_n^+ & \text{ for } t_n < x < t_{n +1}, \ n \in \{1, 2, \dots N - 1\}, \\
C_{N}^+ & \text{ for } t_{n + 1} < x.
\end{cases}
\end{align}
Furthermore, at the discontinuities at $x = \pm t_n$, we can set $\Psi_\omega(x) = (\Psi_\omega(x^+) + \Psi_\omega(x^-))/2$. Integrating Eq.~\ref{eq:scat_state_tdf_1} across the discontinuities at $x = \pm t_n$ we obtain:
\begin{subequations}
\begin{align}
&C_n^- = 1 + \sum_{m = n+1}^N i\sqrt{\gamma_m}\beta_m(\omega) e^{-i\omega t_m} \ \text{for } n \in \{1, 2 \dots N - 1\}, \\
&C_0 = 1 + \sum_{m=1}^N i\sqrt{\gamma_m}\beta_m(\omega) e^{-i\omega t_m}, \\
&C_n^+ = 1 + \sum_{m = 1}^N i\sqrt{\gamma_m}\beta_m(\omega)e^{-i\omega t_m} - \sum_{m=1}^{n}i\sqrt{\gamma_m} \beta_m(\omega) e^{i\omega t_m} \ \text{for } n \in \{1, 2 \dots N\}.
\end{align}
\end{subequations}
Finally, using Eq.~\ref{eq:scat_state_tdf_2}, we obtain the following system of equations for the coefficients $\beta_1(\omega), \beta_2(\omega) \dots \beta_N(\omega)$:
\begin{align}
\begin{bmatrix}
\omega - \omega_1 + 2\gamma_1\sin(\omega t_1) e^{-i\omega t_1} & 2\sqrt{\gamma_1 \gamma_2}\sin(\omega t_1) e^{-i\omega t_2} & \dots & 2\sqrt{\gamma_1 \gamma_N}\sin(\omega t_1)e^{-i\omega t_N}\\
2\sqrt{\gamma_2 \gamma_1}\sin(\omega t_1) e^{-i\omega t_2} & \omega - \omega_2 + 2\gamma_2 \sin(\omega t_2) e^{-i\omega t_2} & \dots & 2\sqrt{\gamma_2 \gamma_N} \sin(\omega t_2) e^{-i\omega t_N} \\
\vdots & \vdots & \ddots & \vdots \\
2\sqrt{\gamma_N \gamma_1} \sin(\omega t_1) e^{-i\omega t_N} & 2\sqrt{\gamma_N \gamma_2} \sin(\omega t_2) e^{-i\omega t_N} &\dots & \omega - \omega_N + 2\gamma_N \sin(\omega t_N)e^{-i\omega t_N}
\end{bmatrix}
\begin{bmatrix}
\beta_1(\omega) \\
\beta_2(\omega) \\
\vdots \\
\beta_N(\omega)
\end{bmatrix} =\nonumber \\ \begin{bmatrix}
2i\sqrt{\gamma_1}\sin(\omega t_1) \\
2i\sqrt{\gamma_2}\sin(\omega t_2) \\
\vdots\\
2i\sqrt{\gamma_N}\sin(\omega t_N)
\end{bmatrix}.
\end{align}

Next, we consider the bound states for this system. We will restrict ourselves to time-delayed feedback systems where all the emitters are at the same frequency $\omega_1 = \omega_2 \dots = \omega_N = \omega_0$ and the time-delays $t_1, t_2 \dots t_N$ all satisfy $\omega_0 t_k = n_k \pi$ for some integer $n_k$ and for all $k \in \{1, 2 \dots N\}$. Under these conditions, as is shown below, this system supports $N$ bound states all at frequency $\omega_0$. For a bound state at frequency $\omega_0$ and with annihilation operator given by Eq.~\ref{eq:bound_state_ansatz}, Eq.~\ref{eq:bs_eq_1} yields:
\begin{align}\label{eq:bs_diff_eq}
i\frac{\partial \Phi_\alpha(x)}{\partial x} + \sum_{n=1}^N \sqrt{\gamma_n}\big(\delta(x + t_n) - \delta(x - t_n)\big) v_n^\alpha = \omega_0 \Phi_\alpha(x).
\end{align}
Since $\Phi_\alpha(x)$ is 0 as $|x|\to \infty$, the solution to this equation can be written as:
\begin{align}\label{eq:bs_td}
\Phi_\alpha(x) = e^{-i\omega_0 x} \begin{cases}
0 & \text{for } x < -t_N \\
B_n^- & \text{for } -t_{n + 1} < x < t_n, \ n \in \{1, 2 \dots N - 1\}, \\
B_0 & \text{for } -t_1 < x < t_1 \\
B_n^+ & \text{for } t_n < x < t_{n + 1}, \ n \in \{1, 2 \dots N - 1\}, \\
B_N^+ & \text{for }x >  t_{n + 1}.
\end{cases}
\end{align}
It follows from integration of Eq.~\ref{eq:bs_diff_eq} across the discontinuities at $x = \pm t_n$ that
\begin{subequations}\label{eq:bs_coeff}
\begin{align}
&B_n^- = \sum_{m = n+1}^N i\sqrt{\gamma_m} v_m^\alpha e^{-i\omega_0 t_m} \ \text{for } n \in \{1, 2 \dots N - 1\}, \\
&B_0 = \sum_{m = 1}^N i\sqrt{\gamma_m} v_m^\alpha e^{-i\omega_0 t_m}, \\
&B_n^+ = \sum_{m=1}^N i\sqrt{\gamma_m}v_m^\alpha e^{-i\omega_0 t_m} - \sum_{ m = 1}^{n} i\sqrt{\gamma_m}v_m^\alpha e^{i\omega_0 t_m}\ \text{for } n \in \{1, 2 \dots N\}.
\end{align}
\end{subequations}
We note that if $\omega_0 t_k = n_k \pi$, then $B_N^+ = 0$, indicating that $\Phi_\alpha(x) \neq 0$ only if $|x| \leq t_n$. Furthermore, under the assumption that all the emitters have frequency $\omega_0$, Eq.~\ref{eq:bs_eq_2} requires that $\Phi_\alpha(-t_n) = \Phi_\alpha(t_n) \ \text{for } n \in \{1, 2 \dots N\}$. This condition is already satisfied if $\Phi_\alpha(x)$ is given by Eqs.~\ref{eq:bs_td} and \ref{eq:bs_coeff}. Therefore, any choice of $v_1^\alpha, v_2^\alpha \dots v_N^\alpha$ will yield a valid bound state --- we thus obtain $N$ (linearly independent) degenerate bound states at frequency $\omega_0$.
\subsection{Finite-difference time-domain simulations of two-photon scattering}
\noindent\emph{Dynamics}: Numerical simulations of two-photon scattering from the multi-emitter time-delayed feedback system can be done using Finite Difference Time Domain (FDTD) method. We express the Hamiltonian of the system, $H$, using the position domain annihilation operator $s_x$
\begin{align}
H = \sum_{n=1}^N \omega_n \sigma_n^\dagger \sigma_n - i\int_{-\infty}^\infty s_x^\dagger \frac{\partial s_x}{\partial x} dx + \sum_{n = 1}^N \sqrt{\gamma_n}\big[(s_{x = -t_n} - s_{x = t_n})\sigma_n^\dagger + \text{h.c.}\big],
\end{align}
where each emitter is assumed to be a two-level system with a ground state and an excited state, denoted by $\ket{g_n}$ and $\ket{e_n}$ for the $n^\text{th}$ emitter, and $\sigma_n = \ket{g_n}\bra{e_n}$. We first go into a rotating frame with respect to the waveguide Hamiltonian to obtain the interaction-picture Hamiltonian $V(t)$ given by
\begin{align}
V(t) = \sum_{n= 1}^N \omega_n \sigma_n^\dagger \sigma_n + \sum_{n = 1}^N \sqrt{\gamma_n}\big[(s_{x = -t_n - t} - s_{x = t_n - t})\sigma_n^\dagger + \text{h.c.}\big].
\end{align}
Now, within the two-excitation subspace, the interaction-picture state of the system can generally be written as
\begin{align}\label{eq:fdtd_ansatz}
\ket{\psi(t)} = \sum_{n, m = 1}^N \psi_{n, m}(t) \sigma_n^\dagger \sigma_m^\dagger \ket{\text{G}} + \sum_{n=1}^N \int_{-\infty}^\infty \psi_n(x; t)\sigma_n^\dagger s_x^\dagger \ket{\text{G}} + \int_{-\infty}^\infty \int_{-\infty}^\infty \psi(x_1, x_2; t) s_{x_1}^\dagger s_{x_2}^\dagger \ket{\text{G}}dx_1 dx_2,
\end{align}
where we assume that $\psi(x_1, x_2; t) = \psi(x_2, x_1; t)$, $\psi_{n, m}(t) = \psi_{m, n}(t)$ and $\psi_{n, n}(t) = 0$. From Schroedinger's equation, we can then derive a set of differential equations for the amplitudes $\psi(x_1, x_2; t), \psi_n(x; t)$ and $\psi_{n, m}(t)$. For the two-photon amplitude $\psi(x_1, x_2; t)$ we obtain
\begin{subequations}
\begin{align}\label{eq:two_ph_amp_fdtd}
2i \frac{\partial}{\partial t}\psi(x_1, x_2; t) = \sum_{n= 1}^N \sqrt{\gamma_n}\bigg[&\psi_n(x_1; t)\delta(x_2 + t_n + t) + \psi_n(x_2; t) \delta(x_1 + t_n + t)- \nonumber \\ & \psi_n(x_1; t) \delta(x_2 - t_n + t) - \psi_n(x_2; t) \delta(x_1 - t_n + t)\bigg].
\end{align}
For the single-photon amplitudes $\psi_n(x; t)$, we obtain
\begin{align}\label{eq:single_ph_amp_fdtd}
i\frac{\partial}{\partial t}\psi_n(x; t) = &\omega_n \psi_n(x; t) + 2\sqrt{\gamma_n}\big(\psi(-t_n - t, x; t) - \psi(t_n - t, x; t)\big) +\nonumber\\ &\sum_{m=1}^N 2\sqrt{\gamma_m}\psi_{m, n}(t)\big(\delta(x + t_m + t) - \delta(x - t_m + t)\big).
\end{align}
For the excited state amplitude $\psi_{m, n}(t)$, we obtain
\begin{align}\label{eq:ex_state_amp_fdtd}
2i\frac{\partial}{\partial t}\psi_{m, n}(t) =& (\omega_m + \omega_n) \psi_{m, n}(t) + \big(\sqrt{\gamma_n}\psi_m(-t -t_n; t) + \sqrt{\gamma_m}\psi_n(-t - t_m; t)\big) - \nonumber\\ &\big(\sqrt{\gamma_n}\psi_m(-t + t_n; t)+ \sqrt{\gamma_m}\psi_n(-t + t_m; t)\big).
\end{align}
\end{subequations}
Eq.~\ref{eq:two_ph_amp_fdtd} can be integrated from $t = 0$ to $t$ to obtain
\begin{align}\label{eq:two_ph_amp_fdtd_integrated}
\psi(x_1, x_2; t) =\psi(x_1, x_2; 0) - \frac{i}{2}\sum_{n = 1}^N \sqrt{\gamma_n}\bigg[&\psi_n(x_1; -x_2 - t_n)\Theta(0\leq -x_2 - t_n \leq t) + \psi_n(x_2; -x_1 - t_n) \Theta(0\leq -x_1 - t_n \leq t) -\nonumber\\
&\psi_n(x_1; -x_2 + t_n)\Theta(0\leq -x_2 + t_n \leq t) - \psi_n(x_2; -x_1 + t_n)\Theta(0 \leq -x_1 + t_n \leq t)\bigg],
\end{align}
where $\Theta(x_1 \leq x \leq x_2) = 1$ if $x \in (x_1, x_2)$, $1/2$ if $x \in \{x_1, x_2\}$ and 0 otherwise. From Eq.~\ref{eq:two_ph_amp_fdtd_integrated}, it follows that
\begin{subequations}\label{eq:psi_two_ph_eval}
\begin{align}
\psi(-t_n - t, x; t) =& \psi(-t_n - t, x; 0) - \frac{i}{4}\sqrt{\gamma_n}\psi_n(x; t) -\nonumber \\ & \frac{i}{2}\sum_{m=n+1}^N \sqrt{\gamma_m}\psi_m(x; t + t_n - t_m) \Theta(t \geq t_m - t_n)- \nonumber\\ &\frac{i}{2}\sum_{m=1}^N\sqrt{\gamma_m}\bigg[\psi_m(-t_n - t; -x - t_m) \Theta(0\leq -x - t_m \leq t) - \psi_m(-t_n - t; -x + t_n) \Theta(0 \leq -x + t_m \leq t)\bigg],
\end{align}
\begin{align}
\psi(t_n - t, x; t) =& \psi(t_n - t, x; 0) + \frac{i}{4}\sqrt{\gamma_n}\psi_n(x; t) +\nonumber \\ & \frac{i}{2}\sum_{m=1}^{n - 1}\sqrt{\gamma_m}\psi_m(x; t - t_n + t_m) \Theta(t \geq t_n - t_m)-\frac{i}{2}\sum_{m = 1}^N \sqrt{\gamma_m}\psi_m(x; t - t_n - t_m) \Theta(t \geq t_n + t_m)- \nonumber \\ 
&\frac{i}{2}\sum_{m = 1}^N \sqrt{\gamma_m}\bigg[\psi_m(t_n - t; -x + t_m) \Theta(0\leq -x + t_m \leq t) - \psi_m(t_n - t; -x + t_n)\Theta(0\leq -x + t_m \leq t)\bigg].
\end{align}
\end{subequations}
Furthermore, Eq.~\ref{eq:single_ph_amp_fdtd} can be integrated across the $\delta$-function discontinuity to obtain the following boundary condition:
\begin{align}\label{eq:bc}
&\psi_n(-t-t_m, t^+) - \psi_n(-t-t_m, t^-) = -2i \sqrt{\gamma_m} \psi_{m, n}(t) \text{ and }\nonumber \\
&\psi_n(-t + t_m, t^+) - \psi_n(-t + t_m, t^-) = 2i  \sqrt{\gamma_m} \psi_{m, n}(t) \text{ for } n \in \{1, 2 \dots N\}.
\end{align}
These boundary conditions along with Eq.~\ref{eq:ex_state_amp_fdtd} imply
\begin{align}\label{eq:update_psi_mn}
i\frac{\partial}{\partial t}\psi_{m, n}(t) = &(\omega_m + \omega_n) \psi_{m, n}(t) - i(\gamma_m + \gamma_n) \psi_{m, n}(t) + \frac{1}{2}\bigg(\sqrt{\gamma_n}\psi_m(-t - t_n; t^-) + \sqrt{\gamma_m}\psi_n(-t - t_m; t^-)\bigg) + \nonumber \\
&\frac{1}{2}\bigg(\sqrt{\gamma_n}\psi_m(-t + t_n; t^-) + \sqrt{\gamma_m}\psi_n(-t + t_m; t^{-})\bigg).
\end{align}
Eqs.~\ref{eq:ex_state_amp_fdtd} and \ref{eq:update_psi_mn} together with Eqs.~\ref{eq:psi_two_ph_eval} and \ref{eq:bc} can be numerically solved using a finite difference time domain scheme to simulate two-photon scattering from the multi-emitter system.\\ \ \\
\noindent\emph{Extracting bound-state trapping probabilities}: The FDTD simulations allow us to compute the state when expressed on the basis of the waveguide mode and the emitter's excited states. The probability of exciting various bound states can then be computed. To see this, we rewrite the state $\ket{\psi(t)}$ on the basis of bound states and scattering states:
\begin{align}
\ket{\psi(t)} = \sum_{\alpha, \beta = 1}^{N_b} \tilde{\psi}_{\alpha, \beta}(t)\phi_\alpha^\dagger \phi_\beta^\dagger \ket{\text{G}} + \sum_{\alpha = 1}^{N_b} \int_{-\infty}^\infty \tilde{\psi}_\alpha(\omega; t) \phi_\alpha^\dagger \psi_\omega^\dagger \ket{\text{G}} d\omega + \int_{-\infty}^\infty \int_{-\infty}^\infty \tilde{\psi}(\omega_1, \omega_2; t)\psi_{\omega_1}^\dagger \psi_{\omega_2}^\dagger \ket{\text{G}}d\omega_1 d\omega_2
\end{align}
The wQED systems under consideration have no two-excitation bound states and thus $\tilde{\psi}_{\alpha, \beta}(t) \to 0$ as $t \to \infty$. Furthermore, as $t\to \infty$, a wave packet of scattering states should be completely in the waveguide and consequently
\begin{align}
\psi_n(x; t \to \infty) = \lim_{t\to \infty}\bra{\text{G}}s_x \sigma_n \ket{\psi(t)} = \lim_{t\to \infty} \sum_{\alpha = 1}^{N_b} \varepsilon_n^\alpha \int_{-\infty}^\infty \frac{\tilde{\psi}_\alpha(\omega; t \to \infty)}{\sqrt{2\pi}} \Psi_\omega^*(x)d\omega,
\end{align}
where $\psi_n(x; t)$ is defined in Eq.~\ref{eq:fdtd_ansatz}. In the limit of infinite time, we can use the asymptotic form of $\Psi_\omega(x)$ from Eq.~\ref{eq:psi_asymp} to obtain
\begin{align}\label{eq:asy_conv}
\psi_n(x; t \to \infty) =  \sum_{\alpha = 1}^{N_b} \varepsilon_n^\alpha \tilde{\psi}_\alpha(x),
\end{align}
where
\begin{align}
\tilde{\psi}_\alpha(x) = \int_{-\infty}^\infty \frac{\tilde{\psi}_\alpha(\omega; t \to \infty)}{\sqrt{2\pi}} \tau(\omega)e^{i\omega x} d\omega.
\end{align}
Eq.~\ref{eq:asy_conv} can be used to compute $\tilde{\psi}_\alpha(x)$ in terms of $\psi_n(x; t \to \infty)$. The trapping probability of the $\alpha^\text{th}$ bound state is then computed via:
\begin{align}
P_\alpha = \int_{-\infty}^\infty |\tilde{\psi}_\alpha(\omega; t\to \infty)|^2 d\omega = \int_{-\infty}^\infty |\tilde{\psi}_\alpha(x)|^2dx.
\end{align}